\documentclass[sigconf,nonacm]{acmart}
\usepackage{multirow}
\usepackage{algorithm}
\usepackage{algpseudocode}
\usepackage{makecell}
\usepackage{subcaption}
\usepackage{stfloats}
\usepackage[utf8]{inputenc}
\usepackage[table]{xcolor}
\usepackage{amsmath}
\usepackage{array}
\usepackage{tikz}
\usetikzlibrary{arrows.meta,positioning,fit,calc,shadows,backgrounds}
\usepackage{xurl}
\newcommand{\code}[1]{{\ttfamily\nolinkurl{#1}}}

\newcommand\dnum{\textup{dnum}}

\begin{document}

\title{End-to-End NTT on FPGA for FHE Acceleration}

\author{Mohamed Allam}
\email{mohamed.allam@bsc.es}
\orcid{0009-0002-0182-1794}
\affiliation{\institution{Barcelona Supercomputing Center (BSC)}
  \city{Barcelona}
  \country{Spain}
}
\affiliation{\institution{Universitat Politècnica de Catalunya (UPC)}
  \city{Barcelona}
  \country{Spain}
}

\author{Thomas Spendlhofer}
\email{thomas.spendlhofer@bsc.es}
\orcid{0009-0006-6968-2283}
\affiliation{\institution{Barcelona Supercomputing Center (BSC)}
  \city{Barcelona}
  \country{Spain}
}

\author{Hugo Sanz-González}
\email{hsanzgon@bsc.es}
\orcid{0009-0000-3203-7850}
\affiliation{\institution{Barcelona Supercomputing Center (BSC)}
  \city{Barcelona}
  \country{Spain}
}

\author{Antonio J. Peña}
\email{antonio.pena@bsc.es}
\orcid{0000-0002-3575-4617}
\affiliation{\institution{Barcelona Supercomputing Center (BSC)}
  \city{Barcelona}
  \country{Spain}
}

\renewcommand{\shortauthors}{Allam et al.}

\begin{abstract}
Fully homomorphic encryption (FHE) enables privacy-preserving computation on encrypted data without decryption. Despite its strong security guarantees, FHE incurs substantial computational and memory overhead.
In FHE-based deep learning inference, on average 63\% of the CPU time is spent on the number theoretic transform (NTT) and its inverse (INTT), which reduce the complexity of polynomial multiplication from $O(N^2)$ to $O(N \log N)$.

Prior FPGA works follow two approaches: accelerating entire FHE schemes, which yields significant speedups but ties them to specific parameters and algorithms, or accelerating only the NTT, which applies across schemes, algorithms, and libraries.
Most FPGA NTT designs require interconnects that grow with parallelism and utilize significant resources, which complicates routing on multi-die FPGAs. Additionally, to the best of our knowledge, end-to-end NTT acceleration within an FHE context has so far been left unexplored.

We design a fully unrolled, delay-commutator streaming NTT/\allowbreak INTT kernel, and floorplan it across the three dies of an AMD Alveo V80.
At 400\,MHz, the kernel transforms a polynomial of $2^{16}$ coefficients every 5.12\,$\mu$s, the highest throughput among FPGA designs at this size.
Offloaded end to end, it is up to 22$\times$ faster than the CPU per transform.
We integrate the kernel into the Lattigo FHE library and evaluate it end to end.
With 128-bit secure parameters, it accelerates core FHE operations, such as key switching and bootstrapping, by 1.8--2.2$\times$, and neural network inference by 1.8--2.1$\times$.
\end{abstract}


\maketitle

\section{Introduction}
\label{sec:introduction}

Fully homomorphic encryption (FHE)~\cite{Gentry2009FHE} is a lattice-based, quantum-safe cryptographic primitive that enables computation on encrypted data without decryption.
It is particularly useful for preserving data privacy in cloud computing, since it enables servers to process sensitive information---such as medical or financial records---while maintaining end-to-end confidentiality.
However, FHE incurs significant memory and computational overhead: depending on the cryptographic parameters, it may require orders of magnitude more memory and computation than the same calculation on unencrypted data.

FHE schemes used in machine learning, primarily CKKS~\cite{ckks17}, encode plaintext data into polynomials over large modular rings, so that encrypted computation reduces to polynomial arithmetic.
Polynomial multiplication in particular is a critical performance bottleneck.
The number theoretic transform (NTT) reduces the complexity of this multiplication from $O(N^2)$ to $O(N \log N)$.
The NTT and its inverse (INTT) dominate the cost of key CKKS operations such as key switching and bootstrapping~\cite{heaan_demystified}.
In our experiments, they consume on average $\sim$63\% of the CPU time in FHE neural network inference.

The NTT is costly because of its large working set, irregular access patterns, and modular arithmetic on large integers.
FPGAs are a strong acceleration fit for the NTT, since these enable pipelined architectures tailored to modular arithmetic and memory layouts that match its access patterns.
State-of-the-art FPGA works~\cite{hera,fast,heap,fab,poseidon} often accelerate an entire FHE scheme; despite significant speedups over CPUs, they are tailored to one implementation of a scheme and hard to adapt to new schemes or algorithms.
Dedicated NTT accelerators instead apply across schemes, algorithms, and libraries.
Most FPGA accelerators propose iterative or partially unrolled architectures~\cite{config-ntt,duongngoc2023area,bootstrapping-capable-ntt,openntt,tpntt,sam,sevenstep,fab,heap,hera}, which operate at high parallelism and utilize significant resources. Given the access patterns of the NTT, they generate significant routing congestion and are challenging to place and route on multi-die FPGAs, often exhausting the limited inter-die wires at high frequencies.
Fully unrolled designs alleviate routing demands by dedicating a pipelined hardware stage to each transform stage. Works that reach high parallelism reorder coefficients among groups of stages through addressed buffers, shufflers or transposes~\cite{kim_bug20,autontt,autonest}, while those built on delay commutators, which reorder with fixed delays and switches only, have been implemented at low parallelism or small polynomial sizes~\cite{proteus,parentt,chiu_parhi_tvlsi25,bsr_tcad26,zaman_tcas26,stelzer_tcas26}.

Across these works, timing closure at high frequency, routing congestion, and multi-die placement are rarely discussed, and, to the best of our knowledge, none evaluates end-to-end NTT offloading within an FHE library or application.

In this work, we address these challenges through a fully unrolled, delay-commutator streaming NTT/INTT kernel, floorplanned across the three dies of an AMD Alveo V80 FPGA, integrated into the Lattigo~\cite{lattigo} FHE library and evaluated on privacy-preserving neural network inference.

In summary, our main contributions are:
\begin{itemize}
    \item A placed-and-routed streaming NTT/\allowbreak INTT kernel for $N{=}2^{16}$ and 62-bit primes on a multi-die FPGA, with the highest frequency and throughput among comparable FPGA designs.
    \item Integration of the kernel into the Lattigo FHE library, and evaluation of end-to-end NTT/INTT offloading, up to $\sim$22$\times$ faster than the CPU per transform, and of core CKKS operations, such as key switching and bootstrapping, with up to $\sim$2$\times$ speedup.
    \item End-to-end evaluation on several FHE neural network inference workloads, attaining speedups of up to $\sim$15$\times$ for NTT/INTT and $\sim$2$\times$ for inference.
\end{itemize}

The rest of the paper is structured as follows. Section~\ref{sec:background} introduces FHE, the RNS-CKKS scheme and the NTT. Section~\ref{sec:methodology} describes the kernel architecture and our design considerations. Section~\ref{sec:evaluation_discussion} presents the experimental setup and evaluation. Section~\ref{sec:related_work} discusses related work, and Section~\ref{sec:conclusions} concludes the paper.

\section{Background}
\label{sec:background}

In this section, we provide the necessary theoretical background on FHE, the RNS-CKKS scheme, and NTT.

\subsection{FHE and RNS-CKKS}
FHE is a cryptographic primitive whose security relies on the hardness of lattice-based assumptions, primarily the learning with errors (LWE) and ring-LWE problems~\cite{regev2009lattices,EC:LyuPeiReg10}, which offer post-quantum security guarantees.
CKKS~\cite{ckks17} is an approximate FHE scheme that encodes vectors of real or complex numbers and supports efficient SIMD computation. A message is represented as a plaintext polynomial whose coefficients approximate the target values under a fixed scaling factor. This polynomial is then encrypted to produce a ciphertext of two polynomials $(a, b)$ under secret key $s$, which carries a small noise for security that grows with each operation.
The ability of CKKS to handle real numbers and perform vectorized linear algebra---such as matrix multiplications, convolutions, dot products, and activation approximations---makes it particularly effective for privacy-preserving deep learning inference~\cite{survey_home}.

In CKKS, encrypted computation is performed as approximate arithmetic over the cyclotomic ring \(R_Q = \mathbb{Z}_Q[X]/(X^N + 1)\), where \(N\) is a power of two
and \(Q\) is a large integer modulus. Homomorphic multiplication of two ciphertexts results in a ciphertext of three polynomials $(a, b, c)$ encrypted under $s^2$ and an increased scale.
The \emph{rescaling} operation restores the scale, while \emph{relinearization} restores the ciphertext to the form $(a, b)$, encrypted under the original key $s$. The latter is attained via \emph{key switching}, which is also needed after rotations.

RNS-CKKS~\cite{ckks-rns} uses the Chinese remainder theorem to store each polynomial as $L{+}1$ residue polynomials, or \emph{limbs}, one per prime of the ciphertext modulus $Q = \prod_{i=0}^{L} q_i$.
Rescaling divides a ciphertext by its last prime, removing one limb; a ciphertext with $\ell{+}1$ limbs is at \emph{level}~$\ell$.
RNS-CKKS is often combined with \emph{hybrid} key switching~\cite{hybrid_ks}, which limits noise growth with an auxiliary modulus $P = \prod_{j} p_j$, extending the ring to $R_{QP}$.
Hybrid key switching splits the limbs into $\dnum$ digits, each extended to $QP$ and transformed with NTTs, so the NTT/INTT work grows with $\dnum$~\cite{ark}.

Once a ciphertext is left with a single limb, \emph{bootstrapping} is required to refresh it.
This process raises the ciphertext to a higher level, enabling further multiplications.
Bootstrapping is the most expensive operation in CKKS, performing many rotations and multiplications, each followed by a key switch~\cite{ckks_bootstrapping,bootstrapping}.

\subsection{NTT}
\label{sec:bg_ntt}

\begin{algorithm}[b]
\caption{Cooley--Tukey $\mathbf{NTT}_\psi$}
\label{alg:CT_psi_tfs}
\begin{algorithmic}[1]
\Procedure{NTT}{$A, N, q, \psi$} \Comment{$\psi$: $2N$-th root of unity}
    \For{$m \gets 1, 2, 4, \dots, N/2$}
        \State $t \gets N/(2m)$
        \For{$i \gets 0 \dots m-1$}
            \State $\zeta \gets \psi^{\text{bit\_rev}(m+i)} \bmod q$
            \For{$j \gets 2it \dots (2i+1)t - 1$}
                \State $u \gets A[j]$,\; $v \gets A[j+t] \cdot \zeta \bmod q$
                \State $A[j] \gets u + v \bmod q$,\; $A[j+t] \gets u - v \bmod q$
            \EndFor
        \EndFor
    \EndFor
    \State \Return $A$
\EndProcedure
\end{algorithmic}
\end{algorithm}

NTT is a fundamental algorithm for efficient polynomial multiplication. It reduces the computational complexity from $O(N^2)$ to $O(N \log N)$, where $N$, a power of two, is the polynomial degree, by operating in a polynomial ring $\mathbb{Z}_q[X] /\allowbreak \Phi(X)$.

If $\Phi(X) = X^N - 1$, the NTT computes a \emph{positively-wrapped convolution} (PWC), using an $N$-th primitive root of unity $\omega$ satisfying $\omega^N \equiv 1 \pmod{q}$. However, PWC requires explicit zero-padding to emulate linear convolution, which is inefficient.
To avoid this, FHE schemes typically work in the negacyclic ring $\mathbb{Z}_q[X]/(X^N + 1)$, where polynomial multiplication is a \emph{negatively-wrapped convolution} (NWC). NWC requires a primitive $2N$-th root of unity $\psi$, such that $\psi^N \equiv -1 \pmod{q}$ and $\psi^2 \equiv \omega \pmod{q}$. NWC may be implemented by scaling the inputs and outputs of a standard NTT: pre-multiplying polynomial coefficients by $\psi^i$ before the forward transform and post-multiplying by $\psi^{-i}$ after the inverse transform.

The pre-scaling and post-scaling operations may be fused into the transform itself by merging the powers of $\psi$ into the twiddle factors, producing the $\mathrm{NTT}_\psi$ variant. The forward and inverse transforms are provided in equations~\eqref{eq:nttpsi-forward} and~\eqref{eq:nttpsi-inverse}, respectively.

\begin{equation}
\label{eq:nttpsi-forward}
\widehat{a}_j = \sum_{i=0}^{N-1} a_i\, \psi^{i(2j+1)} \pmod{q}
\end{equation}
\begin{equation}
\label{eq:nttpsi-inverse}
a_i = N^{-1} \sum_{j=0}^{N-1} \widehat{a}_j\, \psi^{-i(2j+1)} \pmod{q}
\end{equation}

This formulation is algebraically equivalent to NWC but performs no scaling outside the transform.

The main difference between the NTT and INTT lies in the type of butterfly operation performed, as well as in the fact
that the INTT requires a final scaling by $N^{-1} \bmod q$.
Forward NTT uses the Cooley--Tukey (CT) butterfly, whereas INTT uses the Gentleman--Sande (GS) butterfly, as shown in equations~\eqref{eq:ct_butterfly} and~\eqref{eq:gs_butterfly}, where $\zeta$ is a power of $\psi$ (the twiddle factor).

\begin{equation}
\begin{aligned}
u' &= u + v \cdot \zeta \pmod{q}, \\
v' &= u - v \cdot \zeta \pmod{q}
\end{aligned}
\label{eq:ct_butterfly}
\end{equation}

\begin{equation}
\begin{aligned}
u' &= u + v \pmod{q}, \\
v' &= (u - v) \cdot \zeta^{-1} \pmod{q}
\end{aligned}
\label{eq:gs_butterfly}
\end{equation}

Algorithm~\ref{alg:CT_psi_tfs} illustrates a CT forward $\mathrm{NTT}_\psi$ with input in natural order and output in bit-reversed order.

\section{Methodology}
\label{sec:methodology}

In this section, we describe our kernel architecture and its main components: the stream interface, the streaming pipeline, the butterfly unit and on-the-fly twiddle factor generation, and finally how we floorplan it across the FPGA dies.

\subsection{Overall Architecture}

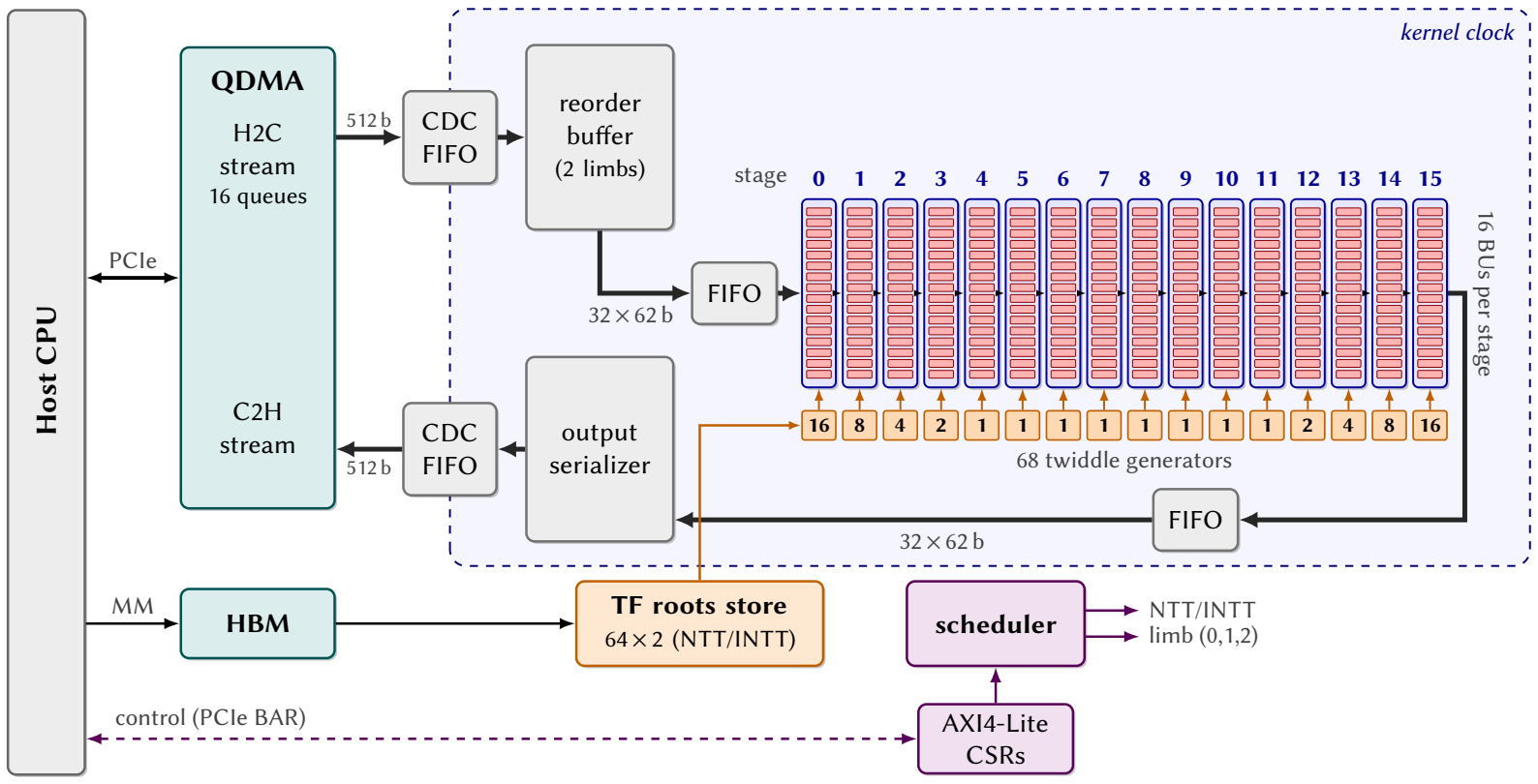
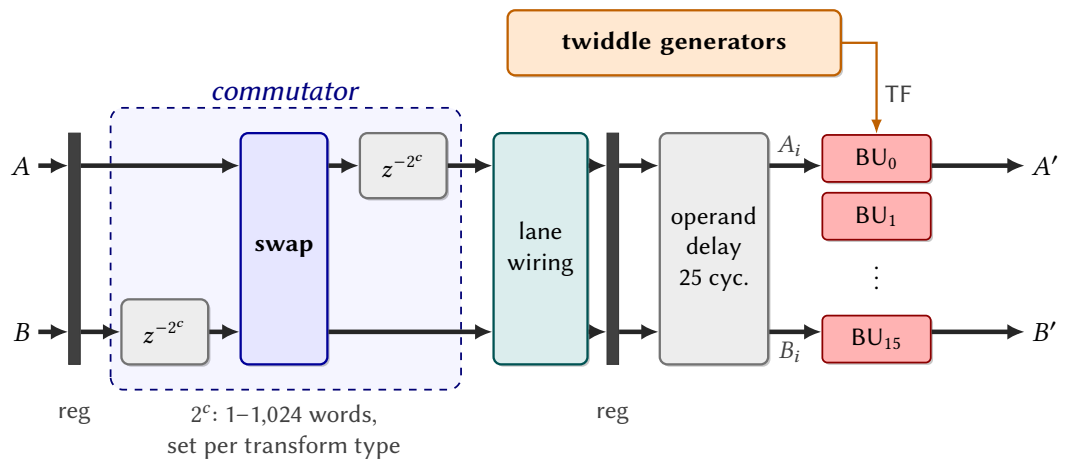
\begin{figure*}[t]
  \centering
  \begin{subfigure}[t]{0.585\textwidth}
  \centering
    \resizebox{\linewidth}{!}{\begin{tikzpicture}[
      font=\sffamily\large,
      sh/.style={drop shadow={shadow xshift=0.9pt, shadow yshift=-0.9pt, opacity=0.22, fill=black}},
      blk/.style={rounded corners=3pt, draw=black!60, line width=0.9pt, align=center, inner sep=3pt, sh},
      mov/.style={blk, fill=teal!14, draw=teal!65!black},
      fif/.style={blk, fill=gray!14, draw=black!55},
      tab/.style={blk, fill=orange!18, draw=orange!75!black},
      ctl/.style={blk, fill=violet!12, draw=violet!70!black},
      bu/.style={draw=red!55!black, fill=red!30, line width=0.35pt, rounded corners=0.4pt, minimum width=3.2mm, minimum height=1.0mm, inner sep=0pt},
      gn/.style={draw=orange!75!black, fill=orange!30, line width=0.6pt, rounded corners=1pt, inner sep=0pt, font=\sffamily\small\bfseries},
      a/.style={-{Latex[length=2mm]}, line width=0.9pt},
      tw/.style={-{Latex[length=1.6mm]}, line width=0.7pt, orange!75!black},
      cl/.style={{Latex[length=2mm]}-{Latex[length=2mm]}, line width=0.9pt, violet!70!black, dashed},
      bus/.style={-{Latex[length=2.4mm, width=2.2mm]}, line width=1.8pt, black!85},
      lbl/.style={font=\sffamily\normalsize, text=black!75},
    ]
    \fill[blue!4, rounded corners=6pt] (2.5,-3.55) rectangle (16.55,3.7);
    \draw[blue!45!black, dashed, line width=0.7pt, rounded corners=6pt] (2.5,-3.55) rectangle (16.55,3.7);
    \node[anchor=north east, font=\sffamily\normalsize\itshape, text=blue!50!black] at (16.45,3.63) {kernel clock};
    \node[blk, fill=black!7, minimum width=10mm, minimum height=99.5mm, anchor=north] (host) at (-2.75,3.7) {};
    \node[rotate=90, font=\sffamily\Large\bfseries] at (-2.75,-1.0) {Host CPU};
    \node[mov, minimum width=20mm, minimum height=60mm] (dma) at (0,0.2) {};
    \node[font=\sffamily\Large\bfseries] at (0,2.75) {QDMA};
    \node[align=center] at (0,1.65) {H2C\\stream\\[-1pt]{\normalsize 16 queues}};
    \node[align=center] at (0,-1.75) {C2H\\stream};
    \node[mov, minimum width=20mm, minimum height=9mm] (hbm) at (0,-4.3) {\bfseries HBM};
    \draw[{Latex[length=2mm]}-{Latex[length=2mm]}, line width=1.4pt] (host.east |- 0,0.2) -- node[above, lbl] {PCIe} (dma.west |- 0,0.2);
    \draw[a] (host.east |- hbm) -- node[above, lbl] {MM} (hbm.west);
    \node[fif, text width=10mm, minimum height=12mm] (hin)  at (2.5, 2.03) {CDC\\FIFO};
    \node[fif, text width=10mm, minimum height=12mm] (hout) at (2.5,-2.03) {CDC\\FIFO};
    \node[fif, minimum height=24mm, text width=17mm] (ct)  at (4.45, 2.03) {reorder\\buffer\\{\normalsize(2 limbs)}};
    \node[fif, minimum height=24mm, text width=17mm] (tap) at (4.45,-2.03) {output\\serializer};
    \draw[bus] (dma.east |- hin) -- node[above, lbl, font=\sffamily\small] {512\,b} (hin);
    \draw[bus] (hout) -- node[below, lbl, font=\sffamily\small] {512\,b} (dma.east |- hout);
    \draw[bus] (hin) -- (ct);
    \draw[bus] (tap) -- (hout);
    \def\X{7.3} \def\DX{0.53}
    \foreach \s/\g in {0/16, 1/8, 2/4, 3/2, 4/1, 5/1, 6/1, 7/1,
                       8/1, 9/1, 10/1, 11/1, 12/2, 13/4, 14/8, 15/16} {
      \pgfmathsetmacro{\x}{\X+\s*\DX}
      \node[rounded corners=2pt, draw=blue!60!black, fill=blue!10, line width=0.7pt,
            minimum width=4.4mm, minimum height=24.5mm, inner sep=0pt, sh] (st\s) at (\x,0) {};
      \foreach \k in {0,...,15} { \node[bu] at (\x,1.06-\k*0.141) {}; }
      \node[font=\sffamily\normalsize\bfseries, text=blue!55!black] at (\x,1.5) {\s};
      \node[gn, minimum width=4.4mm, minimum height=3.8mm] (g\s) at (\x,-1.72) {\g};
      \draw[tw] (g\s.north) -- (st\s.south);
    }
    \foreach \s [count=\n from 1] in {0,...,14} { \draw[-{Latex[length=1.2mm]}, line width=1.1pt] (st\s.east) -- (st\n.west); }
    \pgfmathsetmacro{\XL}{\X+15*\DX+0.4}
    \node[lbl, anchor=east] at (\X-0.3,1.5) {stage};
    \node[lbl, rotate=-90] at (\XL+0.3,0) {16 BUs per stage};
    \node[lbl] at (\X+7.5*\DX,-2.2) {68 twiddle generators};
    \node[fif, minimum width=11mm, minimum height=8mm] (fin)  at (6.2,0) {FIFO};
    \node[fif, minimum width=11mm, minimum height=8mm] (fout) at (12.2,-2.95) {FIFO};
    \draw[bus] (ct.south) |- (fin.west);
    \draw[bus] (fin.east) -- (st0.west);
    \draw[bus] (st15.east) -- ++(0.22,0) |- (fout.east);
    \draw[bus] (fout.west) -- (tap.east |- fout);
    \node[lbl, below] at (4.85,-0.05) {32\,$\times$\,62\,b};
    \node[lbl, below] at (8.9,-3.0) {32\,$\times$\,62\,b};
    \node[tab, text width=30mm, minimum height=11mm] (ts) at (5.75,-4.3) {\bfseries TF roots store\\{\normalsize\mdseries 64\,$\times$\,2 (NTT/INTT)}};
    \node[ctl, text width=21mm, minimum height=11mm] (sch) at (9.6,-4.3) {\bfseries scheduler};
    \node[ctl, text width=18mm, minimum height=9mm] (csr) at (9.6,-5.8) {AXI4-Lite\\CSRs};
    \draw[cl] (host.east |- csr) -- node[above, lbl, pos=0.15] {control (PCIe BAR)} (csr.west);
    \draw[a] (hbm) -- (ts);
    \draw[a, violet!70!black] (csr) -- (sch);
    \draw[tw, line width=0.9pt] (ts.north) |- (g0.west);
    \draw[a, violet!70!black] (sch.east |- 0,-4.13) -- ++(0.7,0) node[right, lbl] {NTT/INTT};
    \draw[a, violet!70!black] (sch.east |- 0,-4.47) -- ++(0.7,0) node[right, lbl] {limb (0,1,2)};
  \end{tikzpicture}}
 \caption{Kernel overview. H2C/C2H: host-to-card/card-to-host streams; CDC: clock-domain crossing; TF: twiddle factor; BU: butterfly unit. Numbers below the stages: twiddle generators per stage.}
  \label{fig:overall_arch}
  \end{subfigure}\hfill
  \begin{subfigure}[t]{0.395\textwidth}
  \centering
    \resizebox{\linewidth}{!}{\begin{tikzpicture}[
      font=\sffamily\large,
      sh/.style={drop shadow={shadow xshift=0.9pt, shadow yshift=-0.9pt, opacity=0.22, fill=black}},
      blk/.style={rounded corners=3pt, draw=black!60, line width=0.9pt, align=center, inner sep=3pt, sh},
      dly/.style={blk, fill=gray!14, draw=black!55, minimum width=12mm, minimum height=9mm},
      sw/.style={blk, fill=blue!10, draw=blue!60!black, minimum width=12mm, minimum height=32mm},
      lw/.style={blk, fill=teal!14, draw=teal!65!black, minimum width=13mm, minimum height=32mm},
      od/.style={blk, fill=gray!14, draw=black!55, minimum width=15mm, minimum height=32mm},
      gen/.style={blk, fill=orange!18, draw=orange!75!black},
      bu/.style={draw=red!55!black, fill=red!30, line width=0.7pt, rounded corners=2pt, minimum width=15mm, minimum height=6.5mm, inner sep=1pt, sh},
      bus/.style={-{Latex[length=2.4mm, width=2.2mm]}, line width=1.8pt, black!85},
      tw/.style={-{Latex[length=2mm]}, line width=0.9pt, orange!75!black},
      reg/.style={draw=black!80, fill=black!75, minimum width=1.6mm, minimum height=32mm, inner sep=0pt},
      grp/.style={rounded corners=5pt, draw=blue!45!black, dashed, line width=0.7pt, fill=blue!4},
      lbl/.style={font=\sffamily\large, text=black!75},
    ]
    \def\yA{1.15} \def\yB{-1.15}
    \fill[blue!4, rounded corners=5pt] (0.65,-1.95) rectangle (5.5,1.95);
    \draw[blue!45!black, dashed, line width=0.7pt, rounded corners=5pt] (0.65,-1.95) rectangle (5.5,1.95);
    \node[anchor=south, font=\sffamily\LARGE\itshape, text=blue!50!black] at (3.05,1.95) {commutator};
    \node[anchor=east, font=\sffamily\Large\bfseries] (ain) at (-0.35,\yA) {$A$};
    \node[anchor=east, font=\sffamily\Large\bfseries] (bin) at (-0.35,\yB) {$B$};
    \node[reg] (r0) at (0.15,0) {};
    \node[dly] (db) at (1.4,\yB) {$z^{-2^{c}}$};
    \node[sw]  (sw) at (3.05,0) {\bfseries swap};
    \node[dly] (da) at (4.7,\yA) {$z^{-2^{c}}$};
    \node[lw]  (lw) at (6.6,0) {lane\\wiring};
    \node[reg] (r1) at (7.6,0) {};
    \node[od]  (od) at (9.0,0) {operand\\delay\\25 cyc.};
    \node[bu] (bu0)  at (11.25,1.25) {BU$_0$};
    \node[bu] (bu1)  at (11.25,0.45) {BU$_1$};
    \node[font=\sffamily\large] at (11.25,-0.3) {$\vdots$};
    \node[bu] (bu15) at (11.25,-1.25) {BU$_{15}$};
    \node[gen, text width=44mm, minimum height=9mm] (g) at (8.45,2.85) {\bfseries twiddle generators};
    \draw[tw] (g.east) -| node[pos=0.78, right, lbl] {TF} (bu0.north);
    \draw[bus] (ain) -- (r0.west |- ain);
    \draw[bus] (bin) -- (r0.west |- bin);
    \draw[bus] (r0.east |- db) -- (db);
    \draw[bus] (db) -- (sw.west |- db);
    \draw[bus] (r0.east |- da) -- (sw.west |- da);
    \draw[bus] (sw.east |- da) -- (da);
    \draw[bus] (da) -- (lw.west |- da);
    \draw[bus] (sw.east |- db) -- (lw.west |- db);
    \draw[bus] (lw.east |- da) -- (r1.west |- da);
    \draw[bus] (lw.east |- db) -- (r1.west |- db);
    \draw[bus] (r1.east |- da) -- (od.west |- da);
    \draw[bus] (r1.east |- db) -- (od.west |- db);
    \draw[bus] (od.east |- da) -- node[above, lbl, pos=0.4] {$A_i$} (bu0.west |- da);
    \draw[bus] (od.east |- db) -- node[below, lbl, pos=0.4] {$B_i$} (bu15.west |- db);
    \node[anchor=west, font=\sffamily\Large\bfseries] (aout) at (13.3,\yA) {$A'$};
    \node[anchor=west, font=\sffamily\Large\bfseries] (bout) at (13.3,\yB) {$B'$};
    \draw[bus] (bu0.east |- da) -- (aout);
    \draw[bus] (bu15.east |- db) -- (bout);
    \node[lbl] at (0.15,-2.3) {reg};
    \node[lbl] at (7.6,-2.3) {reg};
    \node[lbl, align=center, anchor=north] at (3.05,-2.05) {$2^{c}$: 1--1,024 words,\\set per transform type};
  \end{tikzpicture}}
  \caption{Pipeline stage. $A$/$B$: the two 16-lane paths; $z^{-2^{c}}$: delay of $2^{c}$ words; reg: register; BU: butterfly unit; TF: twiddle factor.}
  \label{fig:stage}
  \end{subfigure}
  \caption{The streaming NTT/INTT kernel.}
  \label{fig:kernel}
\end{figure*}

Figure~\ref{fig:overall_arch} shows the kernel, a fully pipelined streaming NTT/INTT architecture for $N = 2^{16}$ and moduli of up to $k = 62$ bits.

Limbs are streamed between host and kernel over the streaming queues of the QDMA engine, without staging in HBM. HBM holds only the twiddle roots, which a 256-bit AXI master loads into the roots store once, at setup.

The pipeline consists of 16 stages, each computing one stage of the transform. Every stage contains 16 butterfly units (BUs) and between 1 and 16 on-the-fly twiddle generators, 68 in total. A scheduler coordinates the pipeline and keeps up to three limbs in flight at once.

The host controls the kernel through an AXI4-Lite slave. It starts a job with a single register write and then polls for completion while results are streamed from the card to the host.

\subsection{Stream Interface}
\label{sec:stream_iface}

The host streams limbs through the QDMA engine in AXI4-Stream mode, with 16 queues per direction that share a 512-bit stream at 333\,MHz (21.3\,GB/s).
A limb is split into 16 slices of 32\,KB, one per queue.
Two 512-beat clock-domain crossing (CDC) FIFOs, one per direction, pass the streams between the DMA and the kernel clocks.

\paragraph{Reorder buffer.}
The reorder buffer turns 512-bit input beats, eight coefficients in host order, into pipeline words of 32 coefficients from across the limb.
It consists of 8 banks, each four coefficients wide, followed by two layers of 2:1 multiplexers that place the coefficients on the pipeline lanes.
Each beat is written to two banks in one cycle and each word reads every bank once, so neither writes nor reads conflict.
The buffer holds two limbs; hence, one limb is written while the previous feeds the pipeline through a 4,096-word input FIFO.

\paragraph{Output serializer.}
The output serializer turns each pipeline word into four 512-bit beats of eight coefficients, each padded to 64 bits, without reordering; the host places each coefficient at its index.
It is fed by an 8,192-word output FIFO and consists of a 32-word LUTRAM FIFO whose head is read four times, once per beat, through a 4:1 multiplexer.

Table~\ref{tab:datapath_bufs} lists the buffers of the stream interface.

\begin{table}[t]
\centering
\caption{Buffers in the stream interface.}
\label{tab:datapath_bufs}
\small
\setlength{\tabcolsep}{3pt}
\begin{tabular}{@{}llll@{}}
\toprule
\textbf{Buffer} & \textbf{Clock} & \textbf{Memory} & \textbf{Size} \\
\midrule
CDC FIFO (input)       & DMA $\to$ kernel & BRAM            & $512 \times 512$\,b \\
Reorder buffer         & kernel                & URAM (32)       & $8 \times 4{,}096 \times 248$\,b \\
Pipeline FIFO (input)  & kernel                & URAM (28)       & $4{,}096 \times 1{,}984$\,b \\
Pipeline FIFO (output) & kernel                & URAM (56)       & $8{,}192 \times 1{,}984$\,b \\
Output serializer      & kernel                & LUTRAM          & $32 \times 1{,}984$\,b \\
CDC FIFO (output)      & kernel $\to$ DMA & BRAM            & $512 \times 512$\,b \\
\bottomrule
\end{tabular}

\vspace{2pt}
{\footnotesize DMA clock: 333\,MHz; kernel clock: 400\,MHz.}
\end{table}

\subsection{Streaming Pipeline}
\label{sec:pipeline}

The pipeline consumes one \emph{word} of 32 coefficients per cycle, split into two \emph{paths}, $A$ and $B$, of 16 lanes each, so a limb streams as 2,048 words over consecutive cycles.
The reorder buffer places the two operands of each stage-0 pair on the same lane of $A$ and $B$: for the NTT, whose input is in natural order, the first half of the limb on $A$ and the second half on $B$; for the INTT, whose input is in bit-reversed order, the even positions on $A$ and the odd on $B$.

Each stage (Figure~\ref{fig:stage}) comprises 16 butterflies and computes stage~$s$ of the transform, for both transform types. It performs its $N/2$ butterfly operations over 2,048 \emph{iterations}, one word per iteration: it reorders the incoming $A$ and $B$, generates the twiddle factors its butterflies need, and passes $A'$ and $B'$ to the next stage.

Butterfly $i$ is hardwired to lane $i$ of the $A$ and $B$ paths: it always
reads $A[i]$ and $B[i]$ and writes its two results back to them. Stage inputs must already be aligned when they arrive at the BUs. Table~\ref{tab:stages} lists, for every stage of both transform types,
where the two operands of a pair arrive and their \emph{stride}, the
distance among the two operands in the stream. Three cases occur:
\begin{itemize}
  \item \emph{No stride} (stage~0 of both transform types). The operands are placed in
        the same word and lane, on different paths, and the stage needs no
        reordering.
  \item \emph{Stride of $2^{c}$ words} (NTT stages 1--11, INTT stages
        5--15). The operands are placed on the same path and lane, $2^{c}$ words
        apart. A delay commutator aligns them.
  \item \emph{Stride of $2^{l}$ lanes} (NTT stages 12--15, INTT stages
        1--4). The operands are placed in the same word and path, $2^{l}$ lanes
        apart. A fixed lane wiring aligns them.
\end{itemize}

\begin{table}[t]
\centering
\caption{NTT and INTT stages. For each butterfly, the
input position (word, path, lane) at which its two operands arrive, in input order (natural for NTT, bit-reversed for INTT).}
\label{tab:stages}
\small
\setlength{\tabcolsep}{3pt}
\begin{tabular}{@{}clllll@{}}
\toprule
\textbf{Stage} & \textbf{Pairs} & \textbf{Operand 1} & \textbf{Operand 2} & \textbf{Stride} & \textbf{Via} \\
\midrule
\multicolumn{6}{@{}l}{\textbf{NTT}} \\
0      & $a_j,\ a_{j+2^{15}}$   & $(t, A, i)$ & $(t, B, i)$            & none             & -- \\
1--11  & $a_j,\ a_{j+2^{15-s}}$ & $(t, p, i)$ & $(t{+}2^{11-s}, p, i)$ & $2^{11-s}$ words & delay \\
12     & $a_j,\ a_{j+2^{3}}$    & $(t, p, i)$ & $(t, p, i{+}8)$        & 8 lanes          & wiring \\
13     & $a_j,\ a_{j+2^{2}}$    & $(t, p, i)$ & $(t, p, i{+}4)$        & 4 lanes          & wiring \\
14     & $a_j,\ a_{j+2^{1}}$    & $(t, p, i)$ & $(t, p, i{+}2)$        & 2 lanes          & wiring \\
15     & $a_j,\ a_{j+2^{0}}$    & $(t, p, i)$ & $(t, p, i{+}1)$        & 1 lane           & wiring \\
\midrule
\multicolumn{6}{@{}l}{\textbf{INTT}} \\
0      & $a_j,\ a_{j+2^{0}}$    & $(t, A, i)$ & $(t, B, i)$            & none             & -- \\
1      & $a_j,\ a_{j+2^{1}}$    & $(t, p, i)$ & $(t, p, i{+}1)$        & 1 lane           & wiring \\
2      & $a_j,\ a_{j+2^{2}}$    & $(t, p, i)$ & $(t, p, i{+}2)$        & 2 lanes          & wiring \\
3      & $a_j,\ a_{j+2^{3}}$    & $(t, p, i)$ & $(t, p, i{+}4)$        & 4 lanes          & wiring \\
4      & $a_j,\ a_{j+2^{4}}$    & $(t, p, i)$ & $(t, p, i{+}8)$        & 8 lanes          & wiring \\
5--15  & $a_j,\ a_{j+2^{s}}$    & $(t, p, i)$ & $(t{+}2^{s-5}, p, i)$  & $2^{s-5}$ words  & delay \\
\bottomrule
\end{tabular}
\end{table}

\paragraph{Commutator.}
The commutator (Figure~\ref{fig:stage}) is an input register, a delay line
on path $B$, a $2{\times}2$ swap, a delay line on path $A$, and an output
register.
Each delay line delays by $2^{c}$ words and is 995 bits wide: 16
coefficients plus control bits, which travel with
the data.
Since the same physical stage serves both NTT and INTT, each line
is built for the larger of its two delays (Table~\ref{tab:delays}) and its
depth is selected at run time by the transform type.
Depending on depth, a delay line is implemented as shift registers (SRLs) up to 32 words, BRAM from 64 to 512 words, and URAM for the 1,024-word lines of stages 1 and 15. A RAM delay line is 14
BRAM36 or 14 URAM blocks wide. Together, the lines hold 130,304 coefficients
($1.99N$, 8.1\,Mbit), the only large memory in the pipeline.

The swap is two 2:1 multiplexers per lane, one per output, each selecting
between the current $A$ and the delayed $B$: 32 coefficient-wide
multiplexers sharing one select. The
select is registered and replicated 32 times, one copy per multiplexer. In stages where only one transform type uses the
commutator, a further 2:1 multiplexer on each output bypasses it for the
other, leaving only the input register. The latency of the commutator is $2^{c} + 2$ cycles.

\paragraph{Lane wiring.}
The lane wiring is a fixed crossing of the lanes and utilizes no memory. For a
stride of $2^{l}$ lanes, it routes the operand of $A[i{+}2^{l}]$ onto
$B[i]$ in the lanes with bit~$l$ clear, and that of $B[i{-}2^{l}]$ onto
$A[i]$ in the lanes with bit~$l$ set; the other 16 coefficient positions
are untouched. Because the other transform type uses a commutator at the same
stage, each of the 16 rerouted positions implements a coefficient-wide 2:1
multiplexer selected by the transform type. The multiplexers feed the
register that every stage features after its commutator, so the crossing adds no
latency.

\begin{table}[t]
\centering
\caption{Commutator delay per stage/transform. The larger of the two is built.
S: shift registers (SRL), B: BRAM, U: URAM.}
\label{tab:delays}
\small
\setlength{\tabcolsep}{2.2pt}
\begin{tabular}{@{}lcccccccccccccccc@{}}
\toprule
\textbf{Stage} & 0 & 1 & 2 & 3 & 4 & 5 & 6 & 7 & 8 & 9 & 10 & 11 & 12 & 13 & 14 & 15 \\
\midrule
NTT     & -- & 1K & 512 & 256 & 128 & 64 & 32 & 16 & 8 & 4  & 2  & 1  & -- & -- & -- & -- \\
INTT    & -- & -- & --  & --  & --  & 1  & 2  & 4  & 8 & 16 & 32 & 64 & 128 & 256 & 512 & 1K \\
\midrule
\textbf{Built}   & -- & 1K & 512 & 256 & 128 & 64 & 32 & 16 & 8 & 16 & 32 & 64 & 128 & 256 & 512 & 1K \\
\textbf{Storage} & -- & U  & B   & B   & B   & B  & S  & S  & S & S  & S  & B  & B   & B   & B   & U  \\
\bottomrule
\end{tabular}
\end{table}

\paragraph{Twiddle factor generators.}
A stage needs as many generators as the largest
number of distinct twiddle factors its 16 butterflies require in one
iteration, over both transform types.
The twiddle of the pair $a_j, a_{j+d}$,
where $d$ is the stride of the stage in coefficient indices, depends only on $\lfloor j/2d \rfloor$. When
$d \ge 16$, the 16 pairs of an iteration share this value, hence one generator
serves all 16 butterflies; when $d < 16$, these need $16/d$ distinct
twiddles and as many generators. Table~\ref{tab:gens} lists these counts; each stage implements the larger of its NTT and INTT counts, 68 generators in total.

\begin{table}[b]
\centering
\caption{Distinct twiddle factors per iteration, by stage and transform type. The larger of the two at each stage sets the number of on-the-fly twiddle factor generators implemented.}
\label{tab:gens}
\setlength{\tabcolsep}{2.2pt}
\begin{tabular}{@{}lccccccccccccccccc@{}}
\toprule
\textbf{Stage} & 0 & 1 & 2 & 3 & 4 & 5 & 6 & 7 & 8 & 9 & 10 & 11 & 12 & 13 & 14 & 15 & \textbf{Tot.} \\
\midrule
NTT   & 1  & 1 & 1 & 1 & 1 & 1 & 1 & 1 & 1 & 1 & 1 & 1 & 2 & 4 & 8 & 16 & 42 \\
INTT  & 16 & 8 & 4 & 2 & 1 & 1 & 1 & 1 & 1 & 1 & 1 & 1 & 1 & 1 & 1 & 1  & 42 \\
\textbf{Built} & 16 & 8 & 4 & 2 & 1 & 1 & 1 & 1 & 1 & 1 & 1 & 1 & 2 & 4 & 8 & 16 & \textbf{68} \\
\bottomrule
\end{tabular}
\end{table}

\subsection{Butterfly Unit}
\label{sec:bu}

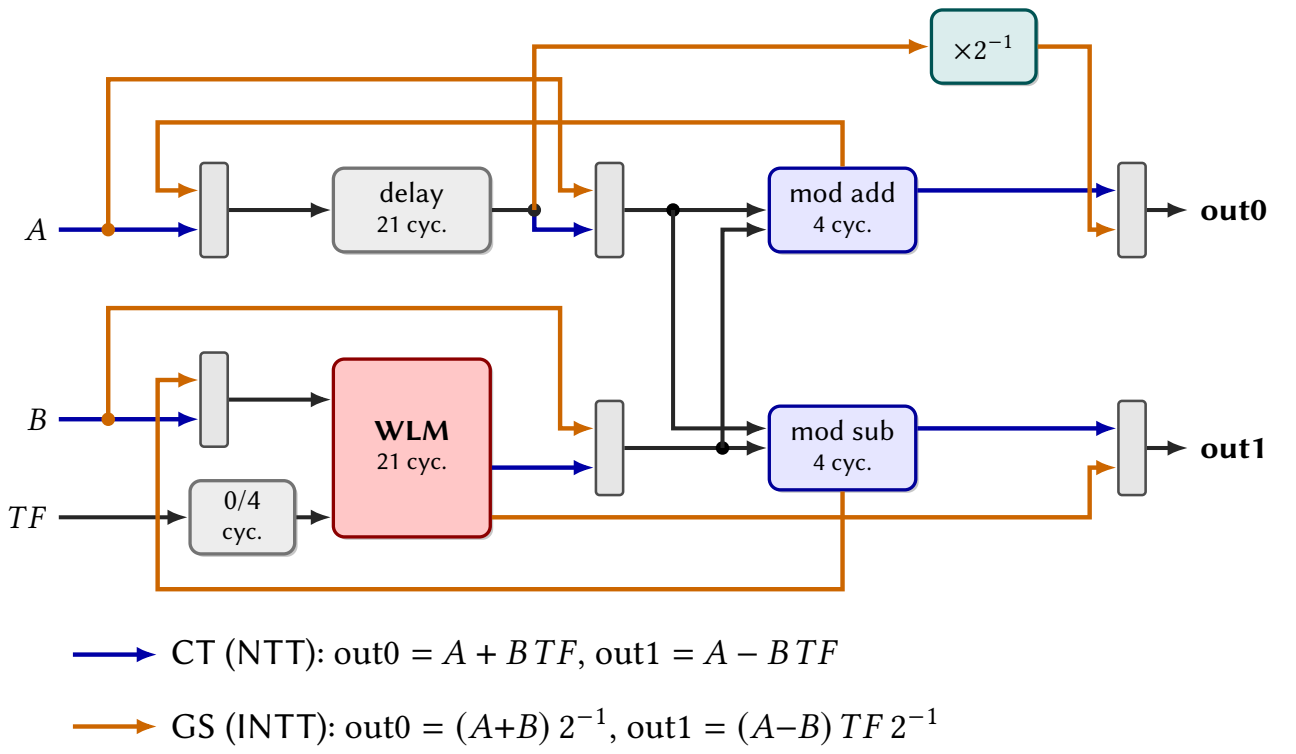
\begin{figure}[t]
  \centering
    \resizebox{\linewidth}{!}{\begin{tikzpicture}[
      x=1.35cm, y=1.25cm,
      font=\sffamily\small,
      sh/.style={drop shadow={shadow xshift=0.8pt, shadow yshift=-0.8pt, opacity=0.22, fill=black}},
      blk/.style={rounded corners=3pt, draw=black!60, line width=0.9pt, align=center, inner sep=2pt, minimum height=8mm, sh},
      mul/.style={blk, fill=red!22, draw=red!55!black, minimum width=15mm},
      ads/.style={blk, fill=blue!10, draw=blue!60!black, minimum width=14mm},
      dly/.style={blk, fill=gray!14, draw=black!55, minimum width=15mm},
      hlf/.style={blk, fill=teal!14, draw=teal!65!black, minimum width=10mm, minimum height=7mm},
      mux/.style={draw=black!70, fill=black!10, line width=0.7pt, rounded corners=1pt, minimum width=2.6mm, minimum height=9mm, inner sep=0pt},
      ct/.style={-{Latex[length=2mm]}, line width=1.2pt, blue!65!black},
      gs/.style={-{Latex[length=2mm]}, line width=1.2pt, orange!80!black},
      both/.style={-{Latex[length=2mm]}, line width=1.1pt, black!85},
      hop/.style={preaction={draw=white, line width=4pt, -}},
      dot/.style={circle, fill, inner sep=1.3pt},
    ]
    \node[anchor=east, font=\sffamily\bfseries] (a)  at (0, 1.05) {$A$};
    \node[anchor=east, font=\sffamily\bfseries] (b)  at (0,-0.4) {$B$};
    \node[anchor=east, font=\sffamily\bfseries] (tf) at (0,-1.15) {$TF$};
    \node[mux] (xd) at (1.1, 1.2) {};
    \node[mux] (xm) at (1.1,-0.25) {};
    \node[dly, minimum width=10mm, minimum height=7mm] (td) at (1.3,-1.15) {0/4\\[-2pt]{\footnotesize cyc.}};
    \node[dly] (d) at (2.5, 1.2) {delay\\[-2pt]{\footnotesize 21 cyc.}};
    \node[mul, minimum height=17mm] (m) at (2.5,-0.62) {\bfseries WLM\\[-2pt]{\footnotesize\mdseries 21 cyc.}};
    \node[mux] (xa) at (3.9, 1.2) {};
    \node[mux] (xb) at (3.9,-0.62) {};
    \node[ads] (ad) at (5.55, 1.2) {mod add\\[-2pt]{\footnotesize 4 cyc.}};
    \node[ads] (sb) at (5.55,-0.62) {mod sub\\[-2pt]{\footnotesize 4 cyc.}};
    \node[hlf] (h) at (6.55, 2.45) {$\times 2^{-1}$};
    \node[mux] (xo0) at (7.6, 1.2) {};
    \node[mux] (xo1) at (7.6,-0.62) {};
    \node[anchor=west, font=\sffamily\bfseries] (o0) at (8.0, 1.2) {out0};
    \node[anchor=west, font=\sffamily\bfseries] (o1) at (8.0,-0.62) {out1};
    \draw[both] (xd.east) -- (d.west);
    \draw[both] (xm.east) -- (m.west |- xm);
    \draw[both] (tf.east) -- (td.west);
    \draw[both] (td.east) -- (m.west |- td);
    \draw[both] (xa.east) -- (ad.west);
    \draw[both] (xb.east) -- (sb.west);
    \node[dot] at (4.35,1.2) {};   \draw[both] (4.35,1.2) -- (4.35,-0.47) -- (sb.west |- 0,-0.47);
    \node[dot] at (4.7,-0.62) {};  \draw[both] (4.7,-0.62) -- (4.7,1.05) -- (ad.west |- 0,1.05);
    \draw[both] (xo0.east) -- (o0);
    \draw[both] (xo1.east) -- (o1);
    \draw[ct] (a.east) -- (xd.west |- a);
    \draw[ct] (b.east) -- (xm.west |- b);
    \coordinate (dj) at ($(d.east)+(0.3,0)$);
    \draw[line width=1.1pt, black!85] (d.east) -- (dj);
    \draw[ct] (dj) |- (xa.west |- 0,1.05);
    \draw[ct] (m.east |- 0,-0.77) -- (xb.west |- 0,-0.77);
    \draw[ct] (ad.east |- 0,1.35) -- (xo0.west |- 0,1.35);
    \draw[ct] (sb.east |- 0,-0.47) -- (xo1.west |- 0,-0.47);
    \node[dot, orange!80!black] at (0.35,1.05) {};
    \draw[gs] (0.35,1.05) -- (0.35,2.2) -- (3.55,2.2) -- (3.55,1.35) -- (xa.west |- 0,1.35);
    \node[dot, orange!80!black] at (0.35,-0.4) {};
    \draw[gs] (0.35,-0.4) -- (0.35,0.45) -| (3.55,-0.47) -- (xb.west |- 0,-0.47);
    \draw[gs] (ad.north) -- (ad.north |- 0,1.87) -- (0.7,1.87) -- (0.7,1.35) -- (xd.west |- 0,1.35);
    \node[dot, black!85] at (dj) {};
    \draw[gs] (dj) -- (dj |- h.west) -- (h.west);
    \draw[gs] (h.east) -- (7.3,2.45) -- (7.3,1.05) -- (xo0.west |- 0,1.05);
    \draw[gs] (m.east |- 0,-1.15) -- (7.3,-1.15) |- (xo1.west |- 0,-0.77);
    \draw[gs] (sb.south) -- (sb.south |- 0,-1.7) -| (0.7,-0.1) -- (xm.west |- 0,-0.1);
    \draw[ct] (0.1,-2.2) -- ++(0.6,0) node[right, black, font=\sffamily\large] {CT (NTT): $\mathrm{out0} = A + B\,TF$, $\mathrm{out1} = A - B\,TF$};
    \draw[gs] (0.1,-2.75) -- ++(0.6,0) node[right, black, font=\sffamily\large] {GS (INTT): $\mathrm{out0} = (A{+}B)\,2^{-1}$, $\mathrm{out1} = (A{-}B)\,TF\,2^{-1}$};
  \end{tikzpicture}}
  \caption{Butterfly unit. WLM: word-level Montgomery multiplier; unlabeled bars: 2:1 multiplexers selected by the transform type.}
  \label{fig:bfly}
\end{figure}

The butterfly unit, in Figure~\ref{fig:bfly}, computes CT butterflies for the forward transform and GS butterflies for the inverse.
Both modes reuse the same units: a modular multiplier, a modular adder, a
modular subtractor and a 21-cycle delay line. Multiplexers set by the
transform type connect them in the CT or the GS order, and both output at
the same latency.

Modular multiplication uses word-level Montgomery reduction (WLM, Algorithm~\ref{alg:wlm}) with $W = 17$-bit words, so that $2^{W} = 2N$; at $k = 62$, this requires $\lambda = 4$ reduction steps.
NTT-friendly primes satisfy $q \equiv 1 \bmod 2N = 2^{W}$. Their Montgomery constant $q' = -q^{-1} \bmod 2^{W}$ is therefore $-1$, and each reduction step needs no multiplication by $q'$. The Montgomery factor $R$ is carried by the twiddle factors, so $\mathrm{WLM}(B, TF)$ returns $B \cdot \psi^{e}$.

\begin{algorithm}[t]
\caption{Word-Level Montgomery multiplication}
\label{alg:wlm}
\begin{algorithmic}
\Procedure{WLM}{$B, TF, q$}
    \State \Comment{$W = 17$, $\lambda = 4$, $R = 2^{W\lambda}$, $q_h = q \gg W$}
    \State $t \gets B \cdot TF$ \Comment{full $2k$-bit product}
    \For{$i \gets 1 \dots \lambda$}
        \State $t_l \gets t \bmod 2^{W}$,\quad $t_h \gets t \gg W$
        \State $t' \gets (-t_l) \bmod 2^{W}$
               \Comment{$q' = -1$: no multiply}
        \State $c \gets t'[W{-}1] \vee t_l[W{-}1]$
               \Comment{$c = 1$ iff $t_l \neq 0$}
        \State $t \gets t_h + q_h \cdot t' + c$
    \EndFor
    \State \textbf{if} $t \ge q$ \textbf{then} $t \gets t - q$
    \State \Return $t$
           \Comment{$= B \cdot TF \cdot R^{-1} \bmod q$}
\EndProcedure
\end{algorithmic}
\end{algorithm}

The inverse transform includes a final scaling by $N^{-1} \bmod q$
(Equation~\eqref{eq:nttpsi-inverse}). Instead of a separate $N^{-1}$ multiplication on the
output, each of the 16 stages scales both GS outputs by $1/2$: together,
these scale by $N^{-1}$. Neither halving costs a multiplier: out0 is halved modulo $q$ by one $k$-bit carry chain (the $\times 2^{-1}$ block in Figure~\ref{fig:bfly}); out1 is halved for free, since $2^{-1} \bmod q$ is folded into the twiddle factors.

The WLM multiplier utilizes 20 DSP58 blocks: 12 for the full $62{\times}62$-bit product and two for each of the four reduction steps. The adder and subtractor are carry chains in LUTs with a conditional correction by $q$, and the delay line utilizes shift registers (SRLs). The 256 BUs utilize 5,120 DSPs.

\subsection{On-the-Fly Twiddle Factor Generation}
\label{sec:twiddle}

The $\mathrm{NTT}_\psi$ requires $N-1$ twiddle factors per prime and transform type (Algorithm~\ref{alg:CT_psi_tfs}).
Stored twiddle factors must be reloaded for every prime of the modulus chain and, since many butterflies share each of these, require either a distribution network, which adds routing congestion, or replicated storage larger than the polynomial itself.
We therefore generate twiddle factors on the fly, trading storage for multipliers.

We split each exponent into its low and high byte, $e = e_{lo} + 256\,e_{hi}$, and keep two sets of 256 \emph{twiddle roots}, $r_{lo}[i] = \psi^{i}$ and $r_{hi}[i] = \psi^{256 i}$ ($2\sqrt{N}$ in total), so that any twiddle factor is one modular multiplication of two roots~\cite{kang13,GPU_NTT_KIM,BTS,zaman_tcas26} (Equation~\eqref{eq:tf_split}).
\begin{equation}
  \psi^{e} = \psi^{e_{lo}} \cdot \big(\psi^{256}\big)^{e_{hi}} = r_{lo}[e_{lo}] \cdot r_{hi}[e_{hi}]
  \label{eq:tf_split}
\end{equation}

\paragraph{TF roots store.}
Since prime chains are known in advance, the host computes the twiddle roots of every prime and transform type once, at setup, and the kernel loads them from HBM into the on-chip \emph{roots store}.
The roots store holds up to 64 primes for both transform types, sufficient for a full CKKS modulus chain, and utilizes 16 URAM.
The host also folds two constants into $r_{lo}$ in advance: the Montgomery factor $R^{2}$ and, for the INTT, the $2^{-1}$ of the GS halving (Section~\ref{sec:bu}).
We denote these precomputed roots $r'_{lo}[i] = \psi^{i}R^{2}$, or $\psi^{i}R^{2}\,2^{-1}$ for the INTT.

\paragraph{Local buffers.}
The generators read the twiddle roots from local buffers in each stage.
Since up to three limbs, each with its own prime, may be in flight, every buffer holds three banks, one per limb; the scheduler fills a bank from the roots store in 512 cycles ahead of its limb, overlapping limb loading.
In every iteration, the generators of a stage share $e_{lo}$ and differ only in $e_{hi}$, so a stage keeps one $r'_{lo}$ buffer for all its generators (three banks of 256 words in two BRAM36), while each generator keeps its own $r_{hi}$ buffer.

\paragraph{On-the-fly twiddle factor generator.}
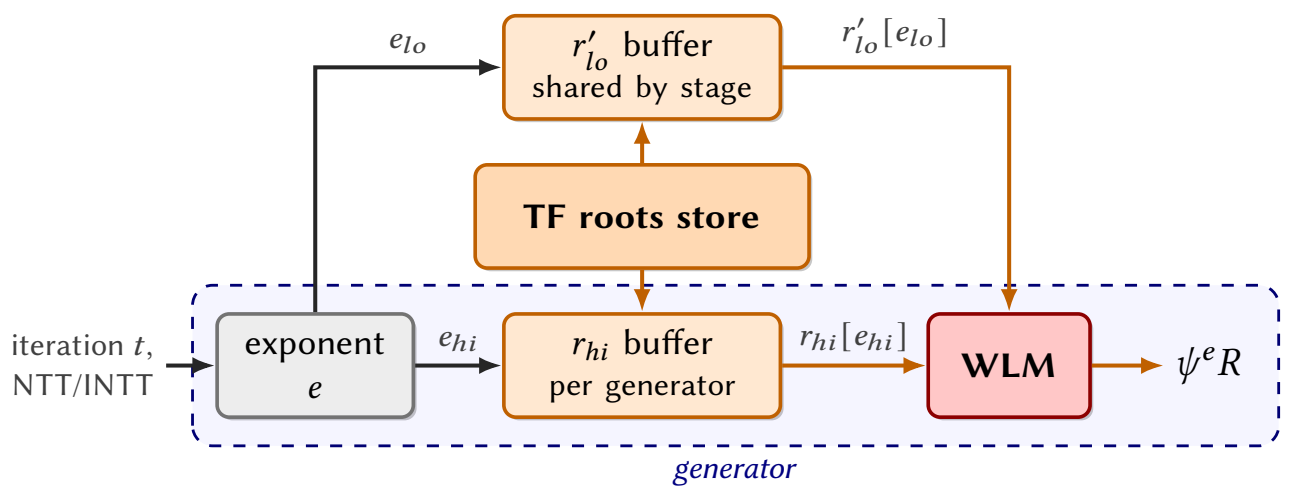
\begin{figure}[b]
  \centering
    \resizebox{\linewidth}{!}{\begin{tikzpicture}[
      font=\sffamily\normalsize,
      sh/.style={drop shadow={shadow xshift=0.8pt, shadow yshift=-0.8pt, opacity=0.22, fill=black}},
      blk/.style={rounded corners=3pt, draw=black!60, line width=0.9pt, align=center, inner sep=3pt, minimum height=9mm, sh},
      mem/.style={blk, fill=orange!18, draw=orange!75!black},
      mul/.style={blk, fill=red!22, draw=red!55!black},
      reg/.style={blk, fill=gray!14, draw=black!55},
      a/.style={-{Latex[length=2mm]}, line width=1.0pt, black!85},
      tw/.style={-{Latex[length=2mm]}, line width=1.0pt, orange!75!black},
      lbl/.style={font=\sffamily\small, text=black!75},
    ]
    \fill[blue!4, rounded corners=5pt] (0.38,-1.55) rectangle (9.85,-0.15);
    \draw[blue!45!black, dashed, line width=0.7pt, rounded corners=5pt] (0.38,-1.55) rectangle (9.85,-0.15);
    \node[anchor=north, font=\sffamily\small\itshape, text=blue!50!black] at (5.1,-1.55) {generator};
    \node[mem, text width=22mm] (rlo) at (4.3, 1.75) {$r'_{lo}$ buffer\\[-1pt]{\small shared by stage}};
    \node[mem, text width=27mm, fill=orange!30] (ts) at (4.3, 0.45) {\bfseries TF roots store};
    \node[lbl, anchor=east, align=right] (in) at (0.15,-0.85) {iteration $t$,\\NTT/INTT};
    \node[reg, text width=15mm] (exp) at (1.45,-0.85) {exponent\\$e$};
    \node[mem, text width=22mm] (rhi) at (4.3,-0.85) {$r_{hi}$ buffer\\[-1pt]{\small per generator}};
    \node[mul, text width=12mm] (m) at (7.5,-0.85) {\bfseries WLM};
    \node[anchor=west, font=\sffamily\large] (out) at (8.85,-0.85) {$\psi^{e}R$};
    \draw[a] (in) -- (exp);
    \draw[a] (exp) -- node[above, lbl] {$e_{hi}$} (rhi);
    \draw[a] (exp.north) |- node[pos=0.75, above, lbl] {$e_{lo}$} (rlo.west);
    \draw[tw] (ts) -- (rlo);
    \draw[tw] (ts) -- (rhi);
    \draw[tw] (rlo.east) -| node[pos=0.25, above, lbl] {$r'_{lo}[e_{lo}]$} (m.north);
    \draw[tw] (rhi) -- node[above, lbl] {$r_{hi}[e_{hi}]$} (m);
    \draw[tw] (m) -- (out);
  \end{tikzpicture}}
  \caption{On-the-fly twiddle factor generator. WLM: word-level Montgomery multiplier.}
  \label{fig:tf_generator}
\end{figure}
The generator (Figure~\ref{fig:tf_generator}) ingests an exponent $e$, derived from the iteration and the transform type, and multiplies its two twiddle roots with one WLM multiplier (Equation~\eqref{eq:tf_gen}). The result is $\psi^{e} R$, the operand the butterfly expects (Section~\ref{sec:bu}).
Each generator utilizes a WLM multiplier (20 DSPs) and its $r_{hi}$ buffer (two BRAM36); the 68 generators utilize 1,360 DSPs and 136 BRAM36.
\begin{equation}
  TF = \mathrm{WLM}\big(r'_{lo}[e_{lo}],\, r_{hi}[e_{hi}]\big) = \psi^{e} R \;(\cdot\,2^{-1}) \bmod q
  \label{eq:tf_gen}
\end{equation}

\subsection{Floorplan}
\label{sec:floorplan}

\begin{figure}[t]
  \centering
    \resizebox{\linewidth}{!}{\begin{tikzpicture}[
      font=\sffamily\large,
      sh/.style={drop shadow={shadow xshift=0.7pt, shadow yshift=-0.7pt, opacity=0.22, fill=black}},
      blk/.style={rounded corners=3pt, draw=black!60, line width=0.8pt, align=center, inner sep=2pt, sh},
      mov/.style={blk, fill=teal!14, draw=teal!65!black, minimum height=9mm},
      tab/.style={blk, fill=orange!18, draw=orange!75!black, minimum height=9mm, font=\sffamily\large\bfseries},
      st/.style={blk, draw=red!55!black, fill=red!22, minimum width=5.4mm, minimum height=6.4mm, inner sep=0pt, font=\sffamily\small\bfseries},
      bus/.style={-{Latex[length=1.4mm]}, line width=1.2pt, black!85},
      tw/.style={-{Latex[length=1.4mm]}, line width=0.8pt, orange!75!black},
      twb/.style={line width=1.4pt, orange!75!black},
      bar/.style={fill=black!75, minimum width=1.2mm, minimum height=5mm, inner sep=0pt},
      tbar/.style={fill=orange!75!black, minimum width=1.2mm, minimum height=5mm, inner sep=0pt},
      slr/.style={rounded corners=5pt, draw=black!45, dashed, line width=0.7pt, fill=black!2},
    ]
    \def\T{3.1} \def\B{0.8} \def\Y{1.95}
    \foreach \k/\x in {0/0.7, 1/5.15, 2/9.6} {
      \draw[slr] (\x,0) rectangle ++(4.3,5.1);
      \node[font=\sffamily\large\bfseries, text=black!70] at (\x+2.15,-0.35) {SLR\k};
    }
    \node[blk, fill=teal!14, draw=teal!65!black, minimum width=5.5mm, minimum height=51mm, anchor=south west] (cpm) at (0,0) {};
    \node[rotate=90, font=\sffamily\large\bfseries] at (cpm) {QDMA};
    \node[blk, fill=teal!14, draw=teal!65!black, minimum width=5.5mm, minimum height=51mm, anchor=south west] (hbm) at (14.05,0) {};
    \node[rotate=90, font=\sffamily\large\bfseries] at (hbm) {HBM};
    \foreach \x in {5.075, 9.525}
      \node[font=\sffamily\normalsize\itshape, text=black!65] at (\x,-0.35) {2 words + TF};
    \foreach \x in {5.075, 9.525} {
      \node[bar] at (\x,\T) {}; \node[bar] at (\x,\B) {}; \node[tbar] at (\x,\Y) {};
    }
    \node[blk, fill=gray!14, draw=black!55, text width=15mm, minimum height=33mm] (si) at (1.65,{(\T+\B)/2}) {stream\\interface};
    \draw[bus] (cpm.east |- 0,\T) -- (si.west |- 0,\T);
    \draw[bus] (si.west |- 0,\B) -- (cpm.east |- 0,\B);
    \foreach \s/\x in {0/3.3, 1/4.3, 2/5.6, 3/6.45, 4/7.3, 5/8.15, 6/9.0,
                       7/10.05, 8/10.9, 9/11.75, 10/12.6}
      \node[st] (s\s) at (\x,\T) {S\s};
    \foreach \s/\x in {11/12.6, 12/8.15, 13/6.45, 14/4.3, 15/3.3}
      \node[st] (s\s) at (\x,\B) {S\s};
    \node[mov, text width=14mm] (axi) at (13.0,4.4) {AXI\\master};
    \node[tab, text width=18mm] (ts)  at (11.0,4.4) {TF roots\\store};
    \draw[twb] (11.325,0 |- ts.south) -- (11.325,\Y);
    \draw[twb] (s0.south |- 0,\Y) -- (s11.north |- 0,\Y);
    \foreach \s in {0,...,10} \draw[tw] (s\s.south |- 0,\Y) -- (s\s.south);
    \foreach \s in {11,...,15} \draw[tw] (s\s.north |- 0,\Y) -- (s\s.north);
    \draw[bus] (si.east |- 0,\T) -- (s0);
    \foreach \s [evaluate=\s as \n using int(\s+1)] in {0,...,9} \draw[bus] (s\s) -- (s\n);
    \draw[bus] (s10.east) -- ++(0.3,0) |- (s11.east);
    \foreach \s [evaluate=\s as \n using int(\s+1)] in {11,...,14} \draw[bus] (s\s) -- (s\n);
    \draw[bus] (s15) -- (si.east |- 0,\B);
    \draw[bus] (hbm.west |- axi) -- (axi);
    \draw[bus] (axi) -- (ts);
  \end{tikzpicture}}
  \caption{Floorplan. S0--S15: pipeline stages; black bars: registered SLR crossings of the stage chain; orange: twiddle-root broadcast from the TF roots store, registered per boundary.}
  \label{fig:floorplan}
\end{figure}
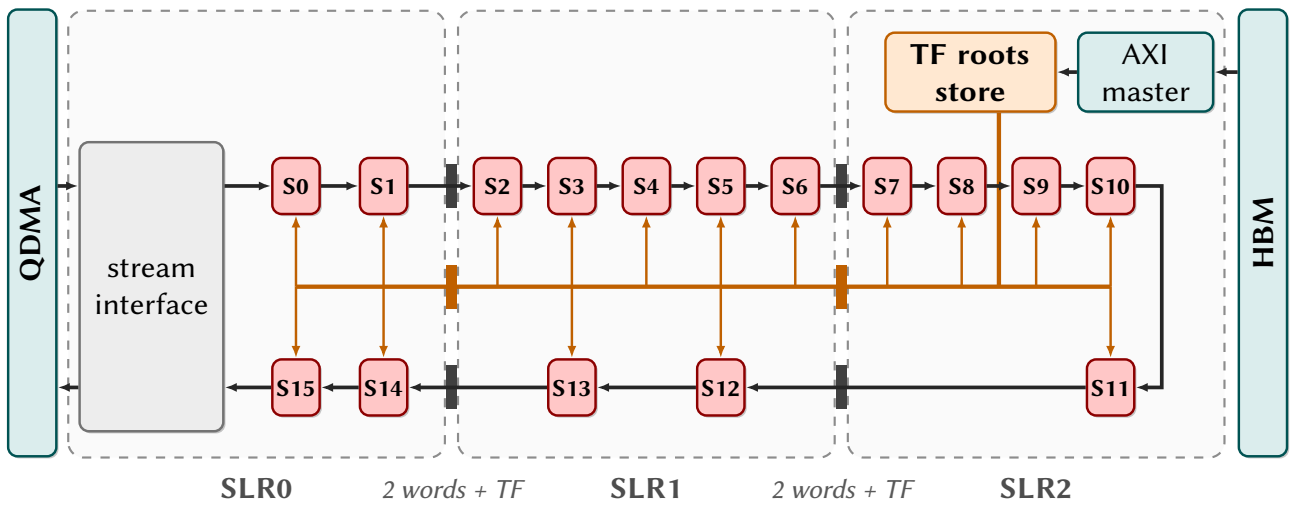
The V80 features three SLRs, with 18,870 super long lines (SLLs) at each boundary. We manually floorplan all kernel cells (Figure~\ref{fig:floorplan}).

The two external interfaces of the kernel are placed at opposite ends of the device: the QDMA engine in SLR0 and the HBM ports next to SLR2.
We therefore place the stream interface in SLR0 and the roots store, with the AXI master that loads it, in SLR2. The AXI master connects to an HBM BLI port, which bypasses the network-on-chip (NoC) of the V80 entirely.
Limbs flow through a U-shaped pipeline that starts and ends next to the stream interface: S0--S1 and S14--S15 in SLR0, S2--S6 and S12--S13 in SLR1, and S7--S11 in SLR2.
Twiddle roots flow against the forward direction of the limbs: for every limb, the roots store broadcasts the modulus and twiddle roots from SLR2 to all stages through a register tree, with one register pair per SLR boundary and one register in front of each stage.
Each boundary is thus traversed only by two stage-to-stage words, one in each direction (S1/S2 and S13/S14, S6/S7 and S11/S12), each 1,984 data bits wide, and by the twiddle-root broadcast.
Each of these paths is registered on both sides of the boundary, and together they utilize fewer than 4,200 SLLs.

Control signals reach the stages through a register tree that limits their fan-out.
Since they change only while the pipeline is idle or once per limb, we also constrain them as two-cycle multicycle paths, which relaxes their routing and eases timing closure: the reset, held for 1,300 cycles; the transform direction, changed only on an empty pipeline; the job configuration registers; and the modulus of each stage, which changes at most once every 2,048 cycles.

\section{Evaluation}
\label{sec:evaluation_discussion}

In this section, we describe the experimental setup, report implementation results, and compare our kernel with state-of-the-art FPGA designs. We then describe the integration into Lattigo, analyze offloading in terms of latency and power, and evaluate it on CKKS operations and FHE neural network inference workloads.

\subsection{Experimental Setup}

Our kernel is written in SystemVerilog and implemented for $N = 2^{16}$ and primes of up to 62 bits on an AMD Alveo V80 card with Vivado 2025.1.
Synthesis and implementation use the default strategies, followed by post-route physical optimization with the \code{Explore} directive.
The card is attached over PCIe Gen5 x8 to a host with two AMD EPYC 9645 processors (96 cores each) and 1.5\,TB of memory.
All CPU baselines use Lattigo~v6.2.0~\cite{lattigo}.
We used an AI assistant to help with analysis and plotting scripts.

\subsection{Implemented Kernel}

\subsubsection{Post-Place-and-Route Results}

The kernel closes timing at 400\,MHz, and we report all results at this frequency.
It utilizes 60\% of the DSPs of the V80 and about a quarter of its LUTs and FFs (Table~\ref{tab:slr_util}). SLR1, which holds seven of the 16 stages, is the most utilized, with 73.5\% of its DSPs.
The design is not limited by routing resources: the kernel shows no routing congestion, and each SLR boundary utilizes 22.1\% of its SLLs.
The frequency is instead bound by wire delay across a clock domain that spans all three SLRs. The critical paths are shallow, 4.65 logic levels on average, but routing accounts for 74\% of their delay.

\begin{table}[b]
\centering
\caption{Kernel per-SLR utilization and SLR crossings.}
\label{tab:slr_util}
\setlength{\tabcolsep}{4pt}
\begin{tabular}{lccc}
\toprule
\textbf{Resource} & \textbf{SLR0} & \textbf{SLR1} & \textbf{SLR2} \\
\midrule
LUT     & 223k (24.8\%) & 269k (32.8\%) & 178k (20.8\%) \\
FF        & 415k (23.0\%) & 505k (30.8\%) & 339k (19.8\%) \\
DSP       & 2,240 (56.2\%) & 2,540 (73.5\%) & 1,700 (49.9\%) \\
BRAM tile & 147 (11.0\%)  & 212 (17.7\%)  & 48 (4.0\%) \\
URAM      & 172 (25.4\%)  & 0             & 16 (2.6\%) \\
\midrule
\textbf{SLLs} & \multicolumn{3}{c}{\textbf{Used / available}} \\
\midrule
SLR0 $\leftrightarrow$ SLR1 & \multicolumn{3}{c}{4,178 / 18,870 (22.1\%): 2,094 up, 2,084 down} \\
SLR1 $\leftrightarrow$ SLR2 & \multicolumn{3}{c}{4,178 / 18,870 (22.1\%): 2,094 up, 2,084 down} \\
\bottomrule
\end{tabular}
\end{table}

\subsubsection{Comparison with State-of-the-Art FPGA Designs}

Table~\ref{tab:comparison} compares the resource utilization, frequency, latency, and throughput of our kernel against prior FPGA NTT implementations.
To weigh performance against hardware cost, we use the area-time product (ATP), the product of the latency and a normalized area term that aggregates LUTs, FFs, DSPs, BRAMs, and URAMs.

\begin{table*}[t]
\centering
\caption{Comparison with state-of-the-art FPGA NTT implementations.}
\label{tab:comparison}
\small
\setlength{\tabcolsep}{0pt}
\renewcommand{\arraystretch}{1.2}
\begin{tabular*}{\textwidth}{@{\extracolsep{\fill}}lcccccccccccc@{}}
\toprule
\textbf{Implementation} & \textbf{Platform} & \textbf{N} & \textbf{log q} & \textbf{Red.} & \textbf{PEs} &
\makecell{\textbf{OTF}\\\textbf{TF}} & \makecell{\textbf{Stored}\\\textbf{TF}} &
\textbf{LUT/FF/BRAM/URAM/DSP} & \makecell{\textbf{Freq.}\\(MHz)} & \makecell{\textbf{Lat.}\\($\mu$s)} &
\makecell{\textbf{Th.put}\\(NTT/s)} & \textbf{ATP} \\
\midrule
Poseidon~\cite{poseidon} & U280 & $2^{16}$ & 32 & Bar & 512 & $\times$ & $N$ &
358k / 344k / 1,024 / 0 / 4,032 & 450 & 80.17 & 12,474 & 92,320 \\
Kurniawan~\cite{config-ntt} & VU37P & $2^{16}$ & 60 & -- & 32 & $\times$ & $N$ &
74.5k / 61.4k / 155$^a$ / 0 / 288 & 250 & 66.51 & 15,035 & 7,702 \\
Mareta~\cite{bootstrapping-capable-ntt} & U250 & $2^{16}$ & 60 & Bar & 16 & \checkmark & 128 &
274.4k / 83.1k / 241$^a$ / 0 / 630 & 250 & 66.2 & 15,106 & 16,266 \\
Duong-Ngoc~\cite{duongngoc2023area} & ZU9EG & $2^{16}$ & 60 & Bar & 32 & \checkmark & 375 &
148.5k / 90.9k / 137$^a$ / 0 / 564 & 200 & 83.8 & 11,933 & 15,761 \\
OpenNTT~\cite{openntt} & ZCU102 & $2^{16}$ & 52 & WLM & 32 & \checkmark & $O(\log N)$ &
157k / 117k / 98 / 0 / 896 & 210 & 78.4 & -- & 19,882 \\
SAM~\cite{sam} & U250 & $2^{16}$ & 64 & -- & -- & P & 304 &
267k / 328k / 2,126$^b$ / -- / 2,736 & 165 & 380 & -- & 414,732 \\
Guo~\emph{et al.}~\cite{guo_tvlsi26} & ZU15EG & $2^{16}$ & 64 & Bar & -- & -- & -- &
82.1k / 76.2k / 356.5$^b$ / -- / 678 & 300 & 54.9 & -- & 13,097 \\
Zaman~\emph{et al.}~\cite{zaman_tcas26} & Virtex-7 & $2^{16}$ & 64 & WLM & 16 & P & 6,206 &
47.3k / 36.3k / 127 / 0 / 399 & 196 & 168.8 & -- & 20,580 \\
Kim~\emph{et al.}~\cite{kim_bug20}$^c$ & VU190 & $2^{17}$ & 62 & Bar & 68 & \checkmark & -- &
365k / 335k / 2,258$^a$ / 0 / 1,332 & 200 & 89.5 & 11,170 & 76,808 \\
Dang Truong~\emph{et al.}~\cite{dangtruong_tvlsi25} & VU13P & $2^{17}$ & 62 & -- & -- & -- & -- &
118.3k / 119.2k / 822 / 0 / 840 & 280 & 257.7 & -- & 97,900 \\
AutoNTT~\cite{autontt}$^d$ & U280 & $2^{16}$ & 64 & WLM & 256 & $\times$ & -- &
503k / 840k / 292 / 128 / 5,120 & 208 & 10.01 & 99,900 & 15,574 \\
AutoNest~\cite{autonest} & U280 & $2^{16}$ & 64 & WLM & 256 & $\times$ & $N$ &
747k / 929k / 176 / 184 / 6,080 & 280 & 8.48 & 117,925 & 15,950 \\
Seven-step~\cite{sevenstep} & U280 & $2^{16}$ & 64 & WLM & 256 & P & $O(\sqrt{N})$ &
460k / 470k / 280 / 0 / 6,320 & 248 & 8.34 & 119,904 & 12,560 \\
TP-NTT~\cite{tpntt} & U280 & $2^{16}$ & 60 & WLM & 256 & $\times$ & $O(N)$ &
361k / 340k / 1,128$^b$ / -- / 2,816 & 250 & 8.28 & -- & 7,692 \\

\textbf{Our Work} & \textbf{V80} & \textbf{$2^{16}$} & \textbf{62} & \textbf{WLM} & \textbf{256} & \textbf{\checkmark} & \textbf{512} &
\textbf{670k / 1,259k / 407 / 188 / 6,480} & \textbf{400} & \textbf{5.12} & \textbf{195.3k} & \textbf{10,567} \\

\bottomrule
\end{tabular*}

\vspace{4pt}
\parbox{\textwidth}{\footnotesize
Red.: WLM = word-level Montgomery, Bar = Barrett. OTF TF: twiddles generated on the fly (P: partial, only inter-dimension twiddles for SAM and Seven-step, the last five stages for Zaman). Stored TF: twiddles stored per prime. Lat.: average time per transform, over consecutive transforms for Zaman, Kim, Seven-step, and TP-NTT, $N/(\text{coefficients per cycle} \cdot f)$ for AutoNTT, AutoNest, and ours, 1/throughput for Poseidon and single-transform latency otherwise. ATP = Lat.\ $\times$ (kLUT + kFF + DSP + BRAM + $8\times$URAM)/5.
$^a$BRAM36 converted from the reported on-chip RAM in KB. $^b$BRAM36 equivalent, including URAM. $^c$INTT. $^d$HLS; reproduced on the U280 in~\cite{autonest}.}
\end{table*}

Designs with lower parallelism~\cite{config-ntt,bootstrapping-capable-ntt,duongngoc2023area,openntt,guo_tvlsi26,zaman_tcas26} run at 196--300\,MHz and require 54.9--168.8\,$\mu$s per transform at $N{=}2^{16}$, 11--33$\times$ our latency, mostly at a higher ATP.
Zaman~\emph{et al.}~\cite{zaman_tcas26}, the only other delay-commutator pipeline in the table, use one butterfly per stage.
Only Kurniawan~\emph{et al.}~\cite{config-ntt} reach a lower ATP, with 288 DSPs, but at 13$\times$ our latency.
At $N{=}2^{17}$, Kim~\emph{et al.}~\cite{kim_bug20} and Dang Truong~\emph{et al.}~\cite{dangtruong_tvlsi25} require 89.5 and 257.7\,$\mu$s, at a higher ATP.
Poseidon~\cite{poseidon} reaches 450\,MHz with 512 butterflies, but on 32-bit moduli, and requires 16$\times$ our latency at a higher ATP; SAM~\cite{sam} requires 74$\times$ our latency, also at a higher ATP.

The closest designs to ours are the four fully pipelined kernels with 256 butterflies on the U280: AutoNTT~\cite{autontt}, AutoNest~\cite{autonest}, Seven-step~\cite{sevenstep}, and TP-NTT~\cite{tpntt}.
These run at 208--280\,MHz and require 8.28--10.01\,$\mu$s per transform, 1.6--2.0$\times$ our latency. AutoNTT, AutoNest and Seven-step experience a higher ATP.
TP-NTT attains a lower ATP, 7,692: it restricts the primes to utilize 11 DSPs per butterfly, against 25 for our kernel, and stores all $N$ twiddle factors per prime, against 512 twiddle roots for ours.

\subsection{Lattigo Integration}
\label{sec:lattigo}
We integrate the kernel into Lattigo~\cite{lattigo}, an open-source FHE library, by offloading its \emph{Ring} transforms, which cover all the limbs of a polynomial, and its \emph{SubRing} transforms, covering a single limb.
We batch the digit decomposition of key switching, which Lattigo transforms one limb at a time, into one call per digit, so that the card amortizes its transfers over multiple limbs.

The offloaded transforms call a C library that we have developed, which interfaces with the QDMA driver and streams the limbs through its queues. Limbs are transferred without staging copies.
The library also loads the twiddle roots of the prime chain at setup.

\subsection{Offloading Analysis}
\label{sec:offload}

\begin{figure}[t]
  \centering
  \includegraphics[width=\linewidth]{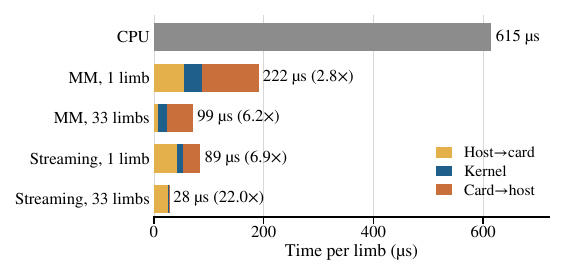}
  \caption{CPU vs.\ kernel offloading in MM mode and streaming mode, 1 and 33 limbs per call, split into host-to-card transfer, kernel, and card-to-host transfer. Labels: end-to-end speedup over CPU.}
  \label{fig:offload_paths}
\end{figure}

\begin{figure}[t]
  \centering
  \includegraphics[width=\linewidth]{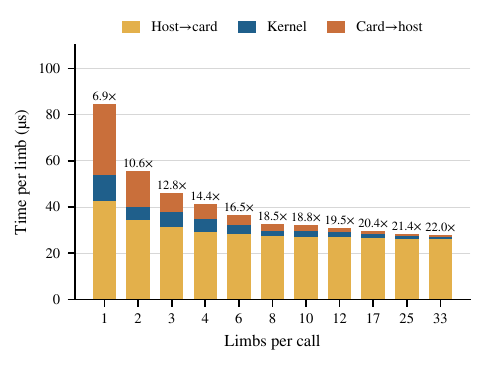}
  \caption{Streaming mode with 1 to 33 limbs per call, time per limb split into host-to-card transfer, kernel, and card-to-host transfer. Labels: end-to-end speedup over CPU.}
  \label{fig:offload_batch}
\end{figure}

We test two offload modes (Figure~\ref{fig:offload_paths}): our streaming mode and memory-mapped (MM) mode, which we implement by instantiating AXI ports on HBM.
In MM mode, the DMA engine copies the limbs into HBM, the kernel reads them and writes the result back, and the DMA engine copies the result to the host.
In this mode, an NTT call of a single non-pipelined limb requires 222\,$\mu$s,
versus 615\,$\mu$s on CPU ($2.8\times$),
but card-to-host transfers dominate it (103\,$\mu$s).
Since the DMA engine must read HBM, these peak at 7.0--8.0\,GB/s, against 15.3--17.2\,GB/s from host to card.
With pipelined calls, the average latency drops to 99\,$\mu$s per limb (6.2$\times$ CPU), still dominated by the card-to-host transfer.

In streaming mode, a call of a single limb requires 89\,$\mu$s (6.9$\times$ CPU): 43\,$\mu$s to stream the limb to the card, 11\,$\mu$s in the kernel, and 31\,$\mu$s to stream the result back.
With 33 pipelined limbs per call, the NTT and the output transfer of each limb overlap the input transfer of the next, and the latency drops to 28\,$\mu$s per limb (22.0$\times$ CPU).

Figure~\ref{fig:offload_batch} shows the per-limb latency of streaming mode with 1 to 33 limbs per call: the speedup over the CPU rises from 6.9$\times$ to 22.0$\times$, since the kernel and the card-to-host transfer are hidden behind the input stream.
From three limbs on, the host-to-card stream accounts for more than 65\% of the time per limb, and at 33 limbs the call transfers at 18.8\,GB/s in each direction, close to the 21.3\,GB/s peak of the stream interface.

\subsection{Power Analysis}
\label{sec:power}

\begin{figure}[t]
  \centering
  \includegraphics[width=\linewidth]{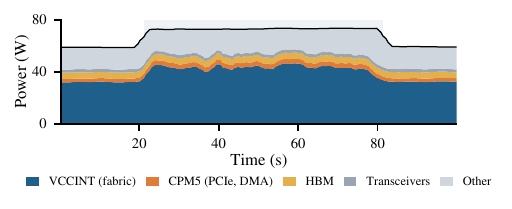}
  \caption{Board power of the V80 by rail, 5-sample moving average. Shaded: 60\,s of NTT executions.}
  \label{fig:power_trace}
\end{figure}

We read the power of the V80 from its board sensors every 0.5\,s while the host runs batches of 10{,}000 pipelined NTTs for 60\,s (Figure~\ref{fig:power_trace}).
The measurement includes the transfers in both directions, since the kernel ingests its input and returns its output over the PCIe streams.
The kernel adds 14.0\,W to the 59.1\,W the card draws while idle, most of it (11.3\,W) in the fabric, while the HBM power stays flat, since no limb passes through HBM.
At one NTT every 26.5\,$\mu$s, the card spends 1.94\,mJ per NTT, 0.37\,mJ above idle.

\subsection{CKKS Operations}
\label{sec:ckks_ops}

\begin{figure}[t]
  \centering
  \begin{subfigure}[b]{0.495\linewidth}
    \centering
    \includegraphics[width=\linewidth]{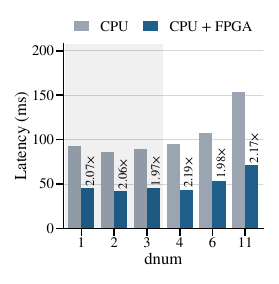}
    \caption{MulRelin, level 10 (11 limbs).}
    \label{fig:mulrelin}
  \end{subfigure}\hfill
  \begin{subfigure}[b]{0.495\linewidth}
    \centering
    \includegraphics[width=\linewidth]{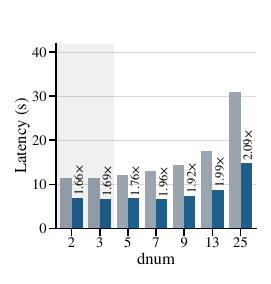}
    \caption{Bootstrapping (25 limbs).}
    \label{fig:bootstrapping}
  \end{subfigure}
  \caption{Latency of MulRelin and bootstrapping, CPU-only vs.\ CPU with NTT/INTT on FPGA, against $\dnum$. Shaded: $\log QP > 1550$, not 128-bit secure~\cite{bootstrapping}. Times are the mean of 10 runs. Labels: end-to-end speedup over CPU.}
  \label{fig:ks_bs_dnum}
\end{figure}

We evaluate the offload on two core CKKS operations: key switching, measured through ciphertext-ciphertext multiplication (MulRelin), and bootstrapping.
The number of NTT calls depends on $\dnum$, the number of digits in the key-switching decomposition~\cite{ark}.
Each digit is extended to the full basis $QP$, which requires transforming its limbs out of and back into the NTT domain, so the number of transforms grows with $\dnum$ (Figure~\ref{fig:ks_bs_dnum}).

For MulRelin at level 10 (Figure~\ref{fig:mulrelin}), the offload is faster at every $\dnum$, from 2.07$\times$ at $\dnum{=}1$ to 2.17$\times$ at $\dnum{=}11$, where it reduces the MulRelin time
from 154.1\,ms to 71.1\,ms.
Within the 128-bit security bound ($\dnum{\ge}4$), the speedup is 1.98--2.19$\times$.

For bootstrapping (Figure~\ref{fig:bootstrapping}), the speedup rises from 1.66$\times$ at $\dnum{=}2$ to 2.09$\times$ at $\dnum{=}25$.
Within the security bound ($\dnum{\ge}5$), it is 1.76--2.09$\times$; at $\dnum{=}5$ ($\log QP = 1546$), the offload reduces the bootstrapping time from 12.04\,s to 6.84\,s.
From $\dnum{=}2$ to 13, the speedup follows the share of CPU time spent in NTT and INTT, which rises from 50.0\% to 63.8\%.

\subsection{FHE Neural Network Inference}
\label{sec:nn_inference}

We evaluate end-to-end inference on six FHE neural networks with Lattigo, following the convolution formulation, polynomial approximation, and bootstrap placement methodology of Orion~\cite{EGR23}.
The networks fall into three categories: a shallow network that requires no bootstrapping; convolutional neural networks (CNNs) that operate on a single ciphertext (ResNet-20 and ResNet-32 on CIFAR-10); and CNNs whose activations span multiple ciphertexts in some layers (AlexNet and VGG16 on CIFAR-10, ResNet-18 on Tiny-ImageNet).
Table~\ref{tab:fhe_parameters} lists the cryptographic parameters, all 128-bit secure~\cite{bootstrapping}.
We exclude setup, which runs once for many inferences.

\begin{figure}[t]
  \centering
  \includegraphics[width=\linewidth]{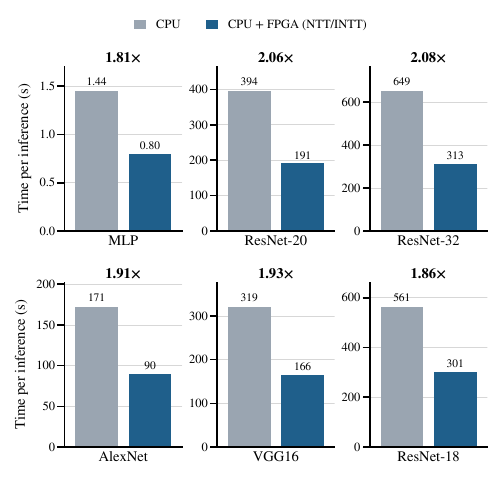}
  \caption{FHE inference time, CPU only vs.\ CPU\,+\,NTT/INTT on the FPGA, mean of 5 inferences. Bar labels: time per inference (s); above each panel: speedup.}
  \label{fig:inference_latency}
\end{figure}

The CPU-only and the offloaded runs produce identical outputs.
Table~\ref{tab:nn_comparison} summarizes the results, and Figure~\ref{fig:inference_latency} compares the time per inference: the offload speeds up inference by 1.81--2.08$\times$.
Per inference, the time Lattigo spends in NTT and INTT falls by 11.4--15.1$\times$ in the shallow and single-ciphertext networks, and by 5.1--14.0$\times$ in the multi-ciphertext networks.

\begin{table}[t]
\centering
\caption{Cryptographic parameters of the workloads.}
\label{tab:fhe_parameters}
\begin{tabular}{@{}lcc@{}}
\toprule
\textbf{Parameter} & \textbf{MLP} & \textbf{CNNs} \\ \midrule
Ring dimension ($N$) & $2^{16}$ & $2^{16}$ \\
Multiplicative depth & 5 & 25$^a$ \\
$\log q_0$ (first prime) & 55 & 55 \\
$\log \Delta$ (scale) & 40 & 40 \\
$\log Q$ & 255 & 1276 \\
$\log P$ & 183 ($3 \times 61$) & 244 ($4 \times 61$) \\
$\log QP$ & 438 & 1520\\
$\dnum$ & 2 & 7 \\
Hamming weight of secret key & 192 & 192 \\ \bottomrule
\end{tabular}

\vspace{2pt}
{\footnotesize $^a$Including bootstrapping levels.\par}
\end{table}

\begin{table}[t]
\centering
\caption{FHE inference: CPU vs.\ CPU\,+\,NTT/INTT on FPGA, per inference (mean of 5 inferences).}
\label{tab:nn_comparison}
\footnotesize
\setlength{\tabcolsep}{0pt}
\renewcommand{\arraystretch}{1.1}
\begin{tabular*}{\columnwidth}{@{\extracolsep{\fill}}lcccccc@{}}
\toprule
                           & \textbf{MLP} & \textbf{ResNet-20} & \textbf{ResNet-32} & \textbf{AlexNet} & \textbf{VGG16} & \textbf{ResNet-18} \\
\midrule
Dataset                    & MNIST & CIFAR-10 & CIFAR-10 & CIFAR-10 & CIFAR-10 & Tiny-IN \\
Bootstraps                 & 0 & 37 & 61 & 15 & 28 & 128 \\
NTT calls                  & 351 & 41.5k & 68.1k & 21.1k & 37.3k & 199k \\
Limbs per call             & 3.81 & 9.52 & 9.54 & 9.08 & 9.22 & 9.11 \\
\midrule
\multicolumn{7}{@{}l}{\textbf{CPU only}} \\
Total (s)                  & 1.44 & 393.9 & 649.1 & 171.4 & 319.3 & 560.7 \\
NTT/INTT (s)               & 0.81 & 238.4 & 392.8 & 105.4 & 192.4 & 430.3 \\
                           & (56.1\%) & (60.5\%) & (60.5\%) & (61.4\%) & (60.1\%) & (76.6\%) \\
\midrule
\multicolumn{7}{@{}l}{\textbf{CPU\,+\,FPGA}} \\
Total (s)                  & 0.80 & 190.8 & 312.7 & 89.6 & 165.8 & 301.0 \\
NTT/INTT (s)               & 0.071 & 15.8 & 26.8 & 7.5 & 15.3 & 84.7 \\
                           & (8.9\%) & (8.3\%) & (8.6\%) & (8.4\%) & (9.2\%) & (28.1\%) \\
Per limb ($\mu$s)          & 52.4 & 39.4 & 40.7 & 38.6 & 44.1 & 43.3 \\
\midrule
\textbf{Speedup}           & \textbf{1.81$\times$} & \textbf{2.07$\times$} & \textbf{2.08$\times$} & \textbf{1.91$\times$} & \textbf{1.93$\times$} & \textbf{1.86$\times$} \\
NTT speedup                & 11.4$\times$ & 15.1$\times$ & 14.7$\times$ & 14.0$\times$ & 12.5$\times$ & 5.1$\times$ \\
\bottomrule
\end{tabular*}
\end{table}

\subsubsection{Shallow Network}

Our multi-layer perceptron (MLP) is a three-layer fully connected network composed of a $784{\times}128$ layer, a $128{\times}128$ layer, and a $128{\times}10$ output layer, with square activations.
Its multiplicative depth is 5, so bootstrapping is not required.
Its short modulus chain limits each call to 3.81 limbs on average, the smallest batches of all workloads; accordingly, the cost per limb is the highest, 52.4\,$\mu$s.
Offloading reduces the time per inference from 1.44\,s to 0.80\,s, a 1.81$\times$ speedup, against an Amdahl bound of 2.28$\times$.

\subsubsection{Single-Ciphertext Networks}
ResNet-20 and ResNet-32 on CIFAR-10 operate on a single ciphertext in every layer.
ResNet-32 is the deeper of the two and requires 61 bootstraps per inference, against 37 for ResNet-20.

Accordingly, ResNet-32 issues 68.1k NTT calls per inference, against 41.5k for ResNet-20.
Their NTT profiles, however, coincide: their calls carry 9.5 limbs on average, and NTT and INTT require 60.5\% of their wall-clock time in the CPU-only runs.

Since these feature similar profiles, the offload accelerates both networks alike.
It reduces the time per inference of ResNet-20 from 393.9\,s to 190.8\,s (2.07$\times$) and that of ResNet-32 from 649.1\,s to 312.7\,s (2.08$\times$), close to the Amdahl bound of 2.53$\times$ set by their NTT share.

\subsubsection{Multi-ciphertext Networks}

AlexNet, VGG16, and ResNet-18 require 15, 28, and 128 bootstraps per inference.
In some of their layers, the activations span several ciphertexts.
Inference then uses multi-threading: each ciphertext is processed by its own thread, and the threads issue their NTT calls concurrently.
AlexNet and VGG16 span two ciphertexts in their first two convolutional layers and a single ciphertext afterward.
ResNet-18, with $64{\times}64{\times}3$ inputs, spans eight, four, two, and one ciphertexts in its four stages.

AlexNet and VGG16 issue 21.1k and 37.3k NTT calls per inference, and ResNet-18 199k.
The calls of all three networks carry 9.1--9.2 limbs on average.
In the CPU-only runs, 1.2, 1.1, and 2.6 threads issue NTT calls at a time on average.
We therefore count the NTT and INTT time as the time during which at least one transform is in flight.
NTT and INTT then require 61.4\% and 60.1\% of the wall-clock time of AlexNet and VGG16, close to the single-ciphertext networks, and 76.6\% in ResNet-18.

AlexNet and VGG16 thus behave like the single-ciphertext networks.
The offload accelerates them by 1.91$\times$ (171.4\,s to 89.6\,s) and 1.93$\times$ (319.3\,s to 165.8\,s), against Amdahl bounds of 2.59$\times$ and 2.51$\times$.
ResNet-18 features a higher bound, 4.27$\times$, but it gains 1.86$\times$ (560.7\,s to 301.0\,s).
With the offload, its concurrent calls queue serially on the card, so its NTT and INTT time falls by only 5.1$\times$, against 14.0$\times$ and 12.5$\times$ for AlexNet and VGG16.

\begin{figure}[t]
  \centering
  \includegraphics[width=\linewidth]{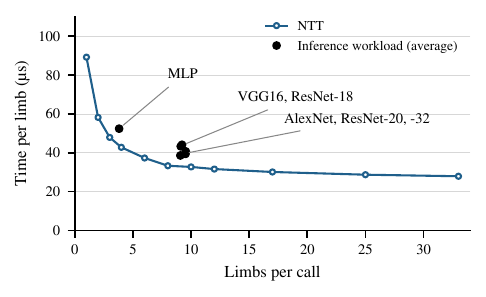}
  \caption{Latency per limb of the inference workloads against the NTT on FPGA: line, NTT batches of 1--33 limbs (Figure~\ref{fig:offload_batch}); dots, latency per offloaded limb in each workload.}

  \label{fig:offload_scaling}
\end{figure}

Across all workloads, the batch size of the calls and multi-threading determine the gain.
At 3.81 limbs per call, the MLP gains 1.81$\times$.
The single-ciphertext networks, at 9.5 limbs per call, gain 2.07--2.08$\times$.
The multi-ciphertext networks issue calls almost as large, 9.1--9.2 limbs, but share the card among concurrent threads and gain 1.86--1.93$\times$.
Figure~\ref{fig:offload_scaling} shows the same trend in the cost per limb: larger calls amortize the fixed cost of each transfer.

\section{Related Work}
\label{sec:related_work}

Hardware acceleration of the NTT has been explored on GPUs, ASICs, and processing-in-memory systems. GPU implementations map the butterflies onto thousands of threads, generate twiddle factors on the fly to relieve memory bandwidth, and fuse kernels onto tensor cores~\cite{GPU_NTT_KIM,GPU_NTT_Ozerk,GPU_NTT_ZHAI,Ozcan2023,Ozcan2022,ISOCC2022,Shivdikar2022,cheddar24,SZWRDM23,warp25,neo}. ASIC accelerators build vectorized NTT units around large on-chip memories, with four-step decompositions, transpose-based or conflict-free memories, and short words that localize communication~\cite{F12021,BTS,craterlake,ark,basalisc,sharp}, and processing-in-memory designs execute the NTT inside DRAM~\cite{BARIK2026108386}.

FPGA accelerators implement either the full scheme or only the NTT. Full-scheme designs~\cite{heax,fab,heap,poseidon,fast,hera} accelerate bootstrapping, but target fixed parameters and specific algorithms. NTT accelerators apply across schemes and libraries. We classify these by how they map the $\log N$ butterfly stages onto hardware and how they reorder coefficients among stages.

Iterative accelerators reuse an array of PEs across stages and reorder the coefficients through banked memory with conflict-free access~\cite{config-ntt,duongngoc2023area,bootstrapping-capable-ntt,config-ntt-arithmatic-core,openntt,guo_tvlsi26,dangtruong_tvlsi25}. These utilize few resources but require 55--84\,$\mu$s per transform at $N{=}2^{16}$.

Dataflow accelerators unroll the stages but still reorder through networks or memories among them. Ye~\emph{et al.}~\cite{ye_isc21} and NTTGen~\cite{nttgen} pair fixed wiring with a permutation network, Kim~\emph{et al.}~\cite{kim_bug20} double-buffer BRAMs among groups of stages, and AutoNTT~\cite{autontt} implements crossbar shufflers on a constant-geometry formulation~\cite{CG}. SAM~\cite{sam}, Seven-step~\cite{sevenstep}, TP-NTT~\cite{tpntt}, and AutoNest~\cite{autonest} decompose the transform into sub-NTTs joined by transposes, and AutoNest splits its transposes into blocks to shorten their critical paths. In both classes, the reordering interconnect grows with parallelism, adding routing congestion and SLR crossings.

Delay commutators avoid this interconnect~\cite{wold_despain84,garrido13}. NTT designs built on these run at low parallelism or small polynomial sizes~\cite{proteus,pipentt,gsmdc,hmntt,parentt,chiu_parhi_tvlsi25,kundi_ashes24,bsr_tcad26,lee_access26}. Stelzer~\emph{et al.}~\cite{stelzer_tcas26} stream up to 64 coefficients per cycle, but only up to $N{=}2^{12}$; Zaman~\emph{et al.}~\cite{zaman_tcas26} reach $N{=}2^{16}$ with one butterfly per stage, and Wang and Ikeda~\cite{wang_ikeda_elex24} stream eight lanes at $N{=}2^{14}$. ASIC accelerators~\cite{osiris,reed,abcfhe} and zero-knowledge accelerators~\cite{pipezk,cyclonentt} adopt such pipelines at larger scale. To share hardware between the NTT and the INTT, these pipelines reverse the data flow~\cite{osiris,kundi_ashes24,stelzer_tcas26}, select different folding sets~\cite{parentt}, or reformulate the data flow~\cite{bsr_tcad26}; Proteus~\cite{proteus} notes that delays sized for both orderings would avoid these changes, but discards the option as too complex. Our kernel streams 32 coefficients per cycle at $N{=}2^{16}$ and computes the NTT and the INTT in the same direction through delays sized for both orderings.

Most works store the twiddle factors, which requires memory that grows with $N$ for every prime~\cite{poseidon,config-ntt,proteus,stelzer_tcas26,lee_access26,tpntt}, and Auto\-Nest fetches those of its last stage from HBM for every transform~\cite{autonest}. Generating every twiddle from $2\sqrt{N}$ twiddle roots with one modular multiplication was introduced in FFT processors~\cite{kang13} and on GPUs~\cite{GPU_NTT_KIM}, and BTS~\cite{BTS} and Osiris~\cite{osiris} dedicate one such multiplier to each PE or butterfly. Zaman~\emph{et al.}~\cite{zaman_tcas26} apply it to the last five stages of a delay-commutator pipeline and store the twiddles of the first eleven. Recurrences need only a few seeds, but one multiplier per PE and a fixed generation order~\cite{openntt,medha,sam,abcfhe,kim_bug20}; powers of a few bases need 375 words and 15 multipliers at $N{=}2^{16}$~\cite{duongngoc2023area}, and power generation needs 254 multipliers~\cite{basalisc}. Our kernel generates all twiddles on chip with 68 generators that share one low-digit buffer per stage and place no constraint on the dataflow.

Most of these works do not discuss placement or multi-die crossings and
none of them evaluates end-to-end offloading within an FHE library. Table~\ref{tab:relwork} summarizes FPGA approaches.

\begin{table}[t]
\centering
\small
\setlength{\tabcolsep}{3pt}
\caption{FPGA FHE and NTT accelerator approaches.}
\label{tab:relwork}
\begin{tabular}{@{}llccl@{}}
\toprule
\textbf{Architecture} & \textbf{Reorder} & \textbf{OTF} & \textbf{E2E} & \textbf{Works} \\
\midrule
PE array      & Memory      & $\times$   & $\times$ & \cite{fab,heap,heax,autontt} \\
PE array      & Memory      & \checkmark & $\times$ & \cite{medha,openntt} \\
PE array      & Network     & $\times$   & $\times$ & \cite{fast} \\
Multi-stage   & Memory      & $\times$   & $\times$ & \cite{poseidon,config-ntt} \\
Multi-stage   & Memory      & \checkmark & $\times$ & \cite{bootstrapping-capable-ntt} \\
Multi-stage   & Network     & $\times$   & $\times$ & \cite{autontt} \\
Sub-NTT       & Network     & $\times$   & $\times$ & \cite{hera} \\
Sub-NTT       & Transpose   & $\times$   & $\times$ & \cite{sam,autonest,sevenstep,tpntt} \\
Unrolled      & Buffers     & \checkmark & $\times$ & \cite{kim_bug20} \\
Unrolled      & Network     & $\times$   & $\times$ & \cite{ye_isc21,autontt,autonest} \\
Unrolled      & Delay       & $\times$   & $\times$ & \cite{proteus,pipentt,gsmdc,parentt,chiu_parhi_tvlsi25,lee_access26,stelzer_tcas26,zaman_tcas26} \\
Unrolled      & Delay       & \checkmark & $\times$ & \cite{wang_ikeda_elex24} \\
\midrule
\textbf{Unrolled} & \textbf{Delay} & \checkmark & \checkmark & \textbf{Our work} \\
\bottomrule
\end{tabular}

\vspace{4pt}
\parbox{\columnwidth}{\footnotesize
PE array: one array iterated over all stages; Multi-stage: a core of several stages iterated over stage groups; Sub-NTT: multi-dimensional decomposition into unrolled sub-NTTs. OTF: all twiddles generated on chip. E2E: placed and routed with per-SLR results, and run end to end under a CPU-hosted FHE library.}
\end{table}

\section{Conclusions}
\label{sec:conclusions}

In this work, we have designed a fully unrolled, delay-commutator streaming NTT/INTT kernel.
Our kernel streams 32 coefficients per cycle through 256 butterfly units in 16 stages and reorders coefficients with delay commutators and fixed lane wiring instead of networks or transposes. A total of 68 generators produce the twiddle factors on the fly from 512 twiddle roots per prime.

We have floorplanned the kernel across the three SLRs of an AMD Alveo V80 and closed timing at 400\,MHz. Our kernel transforms a polynomial of $N{=}2^{16}$ coefficients every 5.12\,$\mu$s.
Offloaded end to end, including the transfers over PCIe, it is up to 22.0$\times$ faster than CPU-only executions.
To the best of our knowledge, among FPGA designs at this polynomial size, it is the delay-commutator NTT with the highest parallelism and attains the highest throughput, 1.6--2.0$\times$ higher than designs with the same parallelism, with a competitive area-time product.

We have integrated the kernel into the Lattigo FHE library and evaluated it on FHE workloads with 128-bit secure parameters.
The offload speeds up MulRelin by 1.98--2.19$\times$, bootstrapping by 1.76--2.09$\times$, and FHE neural network inference by 1.81--2.08$\times$, while the NTT/INTT time per inference falls by up to 15.1$\times$.
The host-to-card stream accounts for more than 65\% of the time per offloaded limb.
Future work may reduce this share with a second stream interface, through the second PCIe controller of the V80, or by integrating the kernel into a larger FHE accelerator to retain limbs on the card.

\begin{acks}
Funded by the European Union (ERC, HomE, 101043467). Views and opinions expressed are however those of the author(s) only and do not necessarily reflect those of the European Union or the European Research Council Executive Agency. Neither the European Union nor the granting authority can be held responsible for them.
Antonio J. Peña was partially supported by the Ramón y Cajal fellowship RYC2020-030054-I, funded by MCIN/AEI/10.13039/501100011033 and by ``ESF Investing in
your future''.
Special thanks to the HomE team, to Antonio Filgueras from OmpSs@FPGA and Juan Miguel De Haro Ruiz from Microarchitecture Design and implementation (MAD) at BSC for the valuable support.
\end{acks}

\bibliographystyle{ACM-Reference-Format}
\bibliography{bibliography}

\begin{thebibliography}{73}

\ifx \showCODEN    \undefined \def \showCODEN     #1{\unskip}     \fi
\ifx \showISBNx    \undefined \def \showISBNx     #1{\unskip}     \fi
\ifx \showISBNxiii \undefined \def \showISBNxiii  #1{\unskip}     \fi
\ifx \showISSN     \undefined \def \showISSN      #1{\unskip}     \fi
\ifx \showLCCN     \undefined \def \showLCCN      #1{\unskip}     \fi
\ifx \shownote     \undefined \def \shownote      #1{#1}          \fi
\ifx \showarticletitle \undefined \def \showarticletitle #1{#1}   \fi
\ifx \showURL      \undefined \def \showURL       {\relax}        \fi
\providecommand\bibfield[2]{#2}
\providecommand\bibinfo[2]{#2}
\providecommand\natexlab[1]{#1}
\providecommand\showeprint[2][]{arXiv:#2}

\bibitem[Aasaraai et~al\mbox{.}(2022)]{cyclonentt}
\bibfield{author}{\bibinfo{person}{Kaveh Aasaraai}, \bibinfo{person}{Emanuele Cesena}, \bibinfo{person}{Rahul Maganti}, \bibinfo{person}{Nicolas Stalder}, \bibinfo{person}{Javier Varela}, {and} \bibinfo{person}{Kevin Bowers}.} \bibinfo{year}{2022}\natexlab{}.
\newblock \bibinfo{title}{{CycloneNTT}: An {NTT/FFT} Architecture Using Quasi-Streaming of Large Datasets on {DDR}- and {HBM}-based {FPGA} Platforms}.
\newblock \bibinfo{howpublished}{Cryptology ePrint Archive, Paper 2022/1657}.
\newblock

\bibitem[Agrawal et~al\mbox{.}(2024)]{heap}
\bibfield{author}{\bibinfo{person}{Rashmi Agrawal}, \bibinfo{person}{Anantha Chandrakasan}, {and} \bibinfo{person}{Ajay Joshi}.} \bibinfo{year}{2024}\natexlab{}.
\newblock \showarticletitle{HEAP: A Fully Homomorphic Encryption Accelerator with Parallelized Bootstrapping}. In \bibinfo{booktitle}{\emph{Proceedings of the 51st Annual International Symposium on Computer Architecture}} (Buenos Aires, Argentina) \emph{(\bibinfo{series}{ISCA '24})}. \bibinfo{publisher}{IEEE Press}, \bibinfo{address}{Buenos Aires, Argentina}, \bibinfo{pages}{756–769}.
\newblock
\showISBNx{9798350326581}
\href{https://doi.org/10.1109/ISCA59077.2024.00060}{doi:\nolinkurl{10.1109/ISCA59077.2024.00060}}

\bibitem[Agrawal et~al\mbox{.}(2023)]{fab}
\bibfield{author}{\bibinfo{person}{Rashmi Agrawal}, \bibinfo{person}{Leo de Castro}, \bibinfo{person}{Guowei Yang}, \bibinfo{person}{Chiraag Juvekar}, \bibinfo{person}{Rabia Yazicigil}, \bibinfo{person}{Anantha Chandrakasan}, \bibinfo{person}{Vinod Vaikuntanathan}, {and} \bibinfo{person}{Ajay Joshi}.} \bibinfo{year}{2023}\natexlab{}.
\newblock \showarticletitle{FAB: An FPGA-Based Accelerator for Bootstrappable Fully Homomorphic Encryption}. In \bibinfo{booktitle}{\emph{2023 IEEE International Symposium on High-Performance Computer Architecture (HPCA)}}. \bibinfo{publisher}{IEEE}, \bibinfo{address}{Montreal, QC, Canada}, \bibinfo{pages}{882--895}.
\newblock
\href{https://doi.org/10.1109/HPCA56546.2023.10070953}{doi:\nolinkurl{10.1109/HPCA56546.2023.10070953}}

\bibitem[Aikata et~al\mbox{.}(2025)]{reed}
\bibfield{author}{\bibinfo{person}{Aikata Aikata}, \bibinfo{person}{Ahmet~Can Mert}, \bibinfo{person}{Sunmin Kwon}, \bibinfo{person}{Maxim Deryabin}, {and} \bibinfo{person}{Sujoy Sinha~Roy}.} \bibinfo{year}{2025}\natexlab{}.
\newblock \showarticletitle{{REED}: Chiplet-based Accelerator for Fully Homomorphic Encryption}.
\newblock \bibinfo{journal}{\emph{IACR Transactions on Cryptographic Hardware and Embedded Systems}} \bibinfo{volume}{2025}, \bibinfo{number}{2} (\bibinfo{year}{2025}), \bibinfo{pages}{163--208}.
\newblock
\href{https://doi.org/10.46586/tches.v2025.i2.163-208}{doi:\nolinkurl{10.46586/tches.v2025.i2.163-208}}

\bibitem[Barik et~al\mbox{.}(2026)]{BARIK2026108386}
\bibfield{author}{\bibinfo{person}{Tathagata Barik}, \bibinfo{person}{Priyam Mehta}, \bibinfo{person}{Zaira Pindado}, \bibinfo{person}{Harshita Gupta}, \bibinfo{person}{Mayank Kabra}, \bibinfo{person}{Mohammad Sadrosadati}, \bibinfo{person}{Onur Mutlu}, {and} \bibinfo{person}{Antonio~J. Peña}.} \bibinfo{year}{2026}\natexlab{}.
\newblock \showarticletitle{Long Integer NTT Execution on UPMEM-PIM for 128-bit Secure Fully Homomorphic Encryption}.
\newblock \bibinfo{journal}{\emph{Future Generation Computer Systems}} (\bibinfo{year}{2026}), \bibinfo{pages}{108386}.
\newblock
\showISSN{0167-739X}
\href{https://doi.org/10.1016/j.future.2026.108386}{doi:\nolinkurl{10.1016/j.future.2026.108386}}

\bibitem[Bossuat et~al\mbox{.}(2021)]{bootstrapping}
\bibfield{author}{\bibinfo{person}{Jean-Philippe Bossuat}, \bibinfo{person}{Christian Mouchet}, \bibinfo{person}{Juan Troncoso-Pastoriza}, {and} \bibinfo{person}{Jean-Pierre Hubaux}.} \bibinfo{year}{2021}\natexlab{}.
\newblock \showarticletitle{Efficient Bootstrapping for Approximate Homomorphic Encryption with Non-Sparse Keys}. In \bibinfo{booktitle}{\emph{Advances in Cryptology -- EUROCRYPT 2021}} \emph{(\bibinfo{series}{Lecture Notes in Computer Science}, Vol.~\bibinfo{volume}{12696})}. \bibinfo{publisher}{Springer}, \bibinfo{address}{Cham}, \bibinfo{pages}{587--617}.
\newblock
\href{https://doi.org/10.1007/978-3-030-77870-5_21}{doi:\nolinkurl{10.1007/978-3-030-77870-5_21}}

\bibitem[Cheon et~al\mbox{.}(2018a)]{ckks_bootstrapping}
\bibfield{author}{\bibinfo{person}{Jung~Hee Cheon}, \bibinfo{person}{Kyoohyung Han}, \bibinfo{person}{Andrey Kim}, \bibinfo{person}{Miran Kim}, {and} \bibinfo{person}{Yongsoo Song}.} \bibinfo{year}{2018}\natexlab{a}.
\newblock \showarticletitle{Bootstrapping for Approximate Homomorphic Encryption}. In \bibinfo{booktitle}{\emph{Advances in Cryptology -- EUROCRYPT 2018}}, \bibfield{editor}{\bibinfo{person}{Jesper~Buus Nielsen} {and} \bibinfo{person}{Vincent Rijmen}} (Eds.). \bibinfo{publisher}{Springer International Publishing}, \bibinfo{address}{Cham}, \bibinfo{pages}{360--384}.
\newblock

\bibitem[Cheon et~al\mbox{.}(2018b)]{ckks-rns}
\bibfield{author}{\bibinfo{person}{Jung~Hee Cheon}, \bibinfo{person}{Kyoohyung Han}, \bibinfo{person}{Andrey Kim}, \bibinfo{person}{Miran Kim}, {and} \bibinfo{person}{Yongsoo Song}.} \bibinfo{year}{2018}\natexlab{b}.
\newblock \showarticletitle{A Full RNS Variant of Approximate Homomorphic Encryption}. In \bibinfo{booktitle}{\emph{Selected Areas in Cryptography – SAC 2018: 25th International Conference, Calgary, AB, Canada, August 15–17, 2018, Revised Selected Papers}} (Calgary, AB, Canada). \bibinfo{publisher}{Springer-Verlag}, \bibinfo{address}{Berlin, Heidelberg}, \bibinfo{pages}{347–368}.
\newblock
\showISBNx{978-3-030-10969-1}
\href{https://doi.org/10.1007/978-3-030-10970-7_16}{doi:\nolinkurl{10.1007/978-3-030-10970-7_16}}

\bibitem[Cheon et~al\mbox{.}(2017)]{ckks17}
\bibfield{author}{\bibinfo{person}{Jung~Hee Cheon}, \bibinfo{person}{Andrey Kim}, \bibinfo{person}{Miran Kim}, {and} \bibinfo{person}{Yongsoo Song}.} \bibinfo{year}{2017}\natexlab{}.
\newblock \showarticletitle{Homomorphic Encryption for Arithmetic of Approximate Numbers}. In \bibinfo{booktitle}{\emph{Advances in Cryptology -- ASIACRYPT 2017}}, \bibfield{editor}{\bibinfo{person}{Tsuyoshi Takagi} {and} \bibinfo{person}{Thomas Peyrin}} (Eds.). \bibinfo{publisher}{Springer International Publishing}, \bibinfo{address}{Cham}, \bibinfo{pages}{409--437}.
\newblock

\bibitem[Chiu and Parhi(2025)]{chiu_parhi_tvlsi25}
\bibfield{author}{\bibinfo{person}{Sin-Wei Chiu} {and} \bibinfo{person}{Keshab~K. Parhi}.} \bibinfo{year}{2025}\natexlab{}.
\newblock \showarticletitle{Architectures for Serial and Parallel Pipelined {NTT}-Based Polynomial Modular Multiplication}.
\newblock \bibinfo{journal}{\emph{IEEE Transactions on Very Large Scale Integration (VLSI) Systems}} \bibinfo{volume}{33}, \bibinfo{number}{9} (\bibinfo{year}{2025}), \bibinfo{pages}{2474--2487}.
\newblock
\href{https://doi.org/10.1109/TVLSI.2025.3576782}{doi:\nolinkurl{10.1109/TVLSI.2025.3576782}}

\bibitem[Dang~Truong et~al\mbox{.}(2025)]{dangtruong_tvlsi25}
\bibfield{author}{\bibinfo{person}{Quang Dang~Truong}, \bibinfo{person}{Phap Duong-Ngoc}, {and} \bibinfo{person}{Hanho Lee}.} \bibinfo{year}{2025}\natexlab{}.
\newblock \showarticletitle{Hybrid Number Theoretic Transform Architecture for Homomorphic Encryption}.
\newblock \bibinfo{journal}{\emph{IEEE Transactions on Very Large Scale Integration (VLSI) Systems}} \bibinfo{volume}{33}, \bibinfo{number}{7} (\bibinfo{year}{2025}), \bibinfo{pages}{2039--2043}.
\newblock
\href{https://doi.org/10.1109/TVLSI.2025.3552852}{doi:\nolinkurl{10.1109/TVLSI.2025.3552852}}

\bibitem[Duong-Ngoc et~al\mbox{.}(2023)]{duongngoc2023area}
\bibfield{author}{\bibinfo{person}{Phap Duong-Ngoc}, \bibinfo{person}{Sunmin Kwon}, \bibinfo{person}{Donghoon Yoo}, {and} \bibinfo{person}{Hanho Lee}.} \bibinfo{year}{2023}\natexlab{}.
\newblock \showarticletitle{Area-Efficient Number Theoretic Transform Architecture for Homomorphic Encryption}.
\newblock \bibinfo{journal}{\emph{IEEE Transactions on Circuits and Systems I: Regular Papers}} \bibinfo{volume}{70}, \bibinfo{number}{3} (\bibinfo{year}{2023}), \bibinfo{pages}{1270--1283}.
\newblock
\href{https://doi.org/10.1109/TCSI.2022.3225208}{doi:\nolinkurl{10.1109/TCSI.2022.3225208}}

\bibitem[Duong-Ngoc et~al\mbox{.}(2022)]{ISOCC2022}
\bibfield{author}{\bibinfo{person}{Phap Duong-Ngoc}, \bibinfo{person}{Thang~Xuan Pham}, \bibinfo{person}{Hanho Lee}, {and} \bibinfo{person}{Tuy~Tan Nguyen}.} \bibinfo{year}{2022}\natexlab{}.
\newblock \showarticletitle{Flexible GPU-Based Implementation of Number Theoretic Transform for Homomorphic Encryption}. In \bibinfo{booktitle}{\emph{2022 19th International SoC Design Conference (ISOCC)}}. \bibinfo{pages}{259--260}.
\newblock

\bibitem[Ebel et~al\mbox{.}(2025)]{EGR23}
\bibfield{author}{\bibinfo{person}{Austin Ebel}, \bibinfo{person}{Karthik Garimella}, {and} \bibinfo{person}{Brandon Reagen}.} \bibinfo{year}{2025}\natexlab{}.
\newblock \showarticletitle{Orion: A Fully Homomorphic Encryption Framework for Deep Learning}. In \bibinfo{booktitle}{\emph{Proceedings of the 30th ACM International Conference on Architectural Support for Programming Languages and Operating Systems, Volume 2}}. \bibinfo{pages}{734–749}.
\newblock

\bibitem[Ebel and Reagen(2026)]{osiris}
\bibfield{author}{\bibinfo{person}{Austin Ebel} {and} \bibinfo{person}{Brandon Reagen}.} \bibinfo{year}{2026}\natexlab{}.
\newblock \showarticletitle{Osiris: A Systolic Approach to Accelerating Fully Homomorphic Encryption}.
\newblock \bibinfo{journal}{\emph{ACM Trans. Archit. Code Optim.}} (\bibinfo{date}{Jan.} \bibinfo{year}{2026}).
\newblock
\showISSN{1544-3566}
\href{https://doi.org/10.1145/3788287}{doi:\nolinkurl{10.1145/3788287}}
\newblock
\shownote{Just Accepted}.

\bibitem[{EPFL-LDS} and {Tune Insight SA}(2024)]{lattigo}
\bibfield{author}{\bibinfo{person}{{EPFL-LDS}} {and} \bibinfo{person}{{Tune Insight SA}}.} \bibinfo{year}{2024}\natexlab{}.
\newblock \bibinfo{title}{Lattigo v6}.
\newblock \bibinfo{howpublished}{Online: \url{https://github.com/tuneinsight/lattigo}}.
\newblock

\bibitem[Fan et~al\mbox{.}(2025)]{warp25}
\bibfield{author}{\bibinfo{person}{G. Fan}, \bibinfo{person}{M. Zhang}, \bibinfo{person}{F. Zheng}, \bibinfo{person}{S. Fan}, \bibinfo{person}{T. Zhou}, \bibinfo{person}{X. Deng}, \bibinfo{person}{W. Tang}, \bibinfo{person}{L. Kong}, \bibinfo{person}{Y. Song}, {and} \bibinfo{person}{S. Yan}.} \bibinfo{year}{2025}\natexlab{}.
\newblock \showarticletitle{WarpDrive: GPU-Based Fully Homomorphic Encryption Acceleration Leveraging Tensor and CUDA Cores}. In \bibinfo{booktitle}{\emph{IEEE Int. Symp. on High Performance Computer Architecture (HPCA)}}. \bibinfo{pages}{1187--1200}.
\newblock

\bibitem[Fan et~al\mbox{.}(2023)]{SZWRDM23}
\bibfield{author}{\bibinfo{person}{Shengyu Fan}, \bibinfo{person}{Zhiwei Wang}, \bibinfo{person}{Weizhi Xu}, \bibinfo{person}{Rui Hou}, \bibinfo{person}{Dan Meng}, {and} \bibinfo{person}{Mingzhe Zhang}.} \bibinfo{year}{2023}\natexlab{}.
\newblock \showarticletitle{TensorFHE: Achieving Practical Computation on Encrypted Data Using GPGPU}. In \bibinfo{booktitle}{\emph{IEEE International Symposium on High-Performance Computer Architecture (HPCA)}}. \bibinfo{pages}{922--934}.
\newblock

\bibitem[Garrido et~al\mbox{.}(2013)]{garrido13}
\bibfield{author}{\bibinfo{person}{Mario Garrido}, \bibinfo{person}{Jes{\'u}s Grajal}, \bibinfo{person}{Miguel~A. S{\'a}nchez}, {and} \bibinfo{person}{Oscar Gustafsson}.} \bibinfo{year}{2013}\natexlab{}.
\newblock \showarticletitle{Pipelined Radix-$2^k$ Feedforward {FFT} Architectures}.
\newblock \bibinfo{journal}{\emph{IEEE Transactions on Very Large Scale Integration (VLSI) Systems}} \bibinfo{volume}{21}, \bibinfo{number}{1} (\bibinfo{year}{2013}), \bibinfo{pages}{23--32}.
\newblock

\bibitem[Geelen et~al\mbox{.}(2022)]{basalisc}
\bibfield{author}{\bibinfo{person}{Robin Geelen}, \bibinfo{person}{Michiel~Van Beirendonck}, \bibinfo{person}{Hilder V.~L. Pereira}, \bibinfo{person}{Brian Huffman}, \bibinfo{person}{Tynan McAuley}, \bibinfo{person}{Ben Selfridge}, \bibinfo{person}{Daniel Wagner}, \bibinfo{person}{Georgios Dimou}, \bibinfo{person}{Ingrid Verbauwhede}, \bibinfo{person}{Frederik Vercauteren}, {and} \bibinfo{person}{David~W. Archer}.} \bibinfo{year}{2022}\natexlab{}.
\newblock \bibinfo{title}{{BASALISC}: Programmable Hardware Accelerator for {BGV} Fully Homomorphic Encryption}.
\newblock \bibinfo{howpublished}{Cryptology {ePrint} Archive, Paper 2022/657}.
\newblock
\urldef\tempurl\url{https://eprint.iacr.org/2022/657}
\showURL{\tempurl}

\bibitem[Geng et~al\mbox{.}(2024)]{gsmdc}
\bibfield{author}{\bibinfo{person}{Yue Geng}, \bibinfo{person}{Xiao Hu}, {and} \bibinfo{person}{Zhongfeng Wang}.} \bibinfo{year}{2024}\natexlab{}.
\newblock \showarticletitle{{GS-MDC}: High-Speed and Area-Efficient Number Theoretic Transform Design}.
\newblock \bibinfo{journal}{\emph{IEEE Transactions on Circuits and Systems II: Express Briefs}} \bibinfo{volume}{71}, \bibinfo{number}{12} (\bibinfo{year}{2024}), \bibinfo{pages}{4974--4978}.
\newblock
\href{https://doi.org/10.1109/TCSII.2024.3430954}{doi:\nolinkurl{10.1109/TCSII.2024.3430954}}

\bibitem[Gentry(2009)]{Gentry2009FHE}
\bibfield{author}{\bibinfo{person}{Craig Gentry}.} \bibinfo{year}{2009}\natexlab{}.
\newblock \emph{\bibinfo{title}{A Fully Homomorphic Encryption Scheme}}.
\newblock \bibinfo{thesistype}{Ph.\,D. Dissertation}. \bibinfo{school}{Stanford University}, \bibinfo{address}{Stanford, CA, USA}.
\newblock

\bibitem[Guo et~al\mbox{.}(2026)]{guo_tvlsi26}
\bibfield{author}{\bibinfo{person}{Jianbo Guo}, \bibinfo{person}{Jiaoyang Zhu}, {and} \bibinfo{person}{Hao Xiao}.} \bibinfo{year}{2026}\natexlab{}.
\newblock \showarticletitle{A High-Performance and Configurable {NTT} Accelerator Based on Scalable {2-D} Architecture}.
\newblock \bibinfo{journal}{\emph{IEEE Transactions on Very Large Scale Integration (VLSI) Systems}} \bibinfo{volume}{34}, \bibinfo{number}{2} (\bibinfo{year}{2026}), \bibinfo{pages}{506--518}.
\newblock
\href{https://doi.org/10.1109/TVLSI.2025.3606636}{doi:\nolinkurl{10.1109/TVLSI.2025.3606636}}

\bibitem[Halevi et~al\mbox{.}(2019)]{hybrid_ks}
\bibfield{author}{\bibinfo{person}{Shai Halevi}, \bibinfo{person}{Yuriy Polyakov}, {and} \bibinfo{person}{Victor Shoup}.} \bibinfo{year}{2019}\natexlab{}.
\newblock \showarticletitle{An Improved RNS Variant of the BFV Homomorphic Encryption Scheme}. In \bibinfo{booktitle}{\emph{Topics in Cryptology -- CT-RSA 2019}}, \bibfield{editor}{\bibinfo{person}{Mitsuru Matsui}} (Ed.). \bibinfo{publisher}{Springer International Publishing}, \bibinfo{pages}{83--105}.
\newblock
\showISBNx{978-3-030-12612-4}

\bibitem[Hirner et~al\mbox{.}(2024)]{proteus}
\bibfield{author}{\bibinfo{person}{Florian Hirner}, \bibinfo{person}{Ahmet~Can Mert}, {and} \bibinfo{person}{Sujoy Sinha~Roy}.} \bibinfo{year}{2024}\natexlab{}.
\newblock \showarticletitle{Proteus: A Pipelined {NTT} Architecture Generator}.
\newblock \bibinfo{journal}{\emph{IEEE Transactions on Very Large Scale Integration (VLSI) Systems}} \bibinfo{volume}{32}, \bibinfo{number}{7} (\bibinfo{year}{2024}), \bibinfo{pages}{1228--1238}.
\newblock
\href{https://doi.org/10.1109/TVLSI.2024.3377366}{doi:\nolinkurl{10.1109/TVLSI.2024.3377366}}

\bibitem[Jiao et~al\mbox{.}(2025)]{neo}
\bibfield{author}{\bibinfo{person}{Dian Jiao}, \bibinfo{person}{Xianglong Deng}, \bibinfo{person}{Zhiwei Wang}, \bibinfo{person}{Shengyu Fan}, \bibinfo{person}{Yi Chen}, \bibinfo{person}{Dan Meng}, \bibinfo{person}{Rui Hou}, {and} \bibinfo{person}{Mingzhe Zhang}.} \bibinfo{year}{2025}\natexlab{}.
\newblock \showarticletitle{Neo: Towards Efficient Fully Homomorphic Encryption Acceleration Using Tensor Core}. In \bibinfo{booktitle}{\emph{52nd Annual International Symposium on Computer Architecture}}. \bibinfo{pages}{107–121}.
\newblock

\bibitem[Jung et~al\mbox{.}(2020)]{heaan_demystified}
\bibfield{author}{\bibinfo{person}{Wonkyung Jung}, \bibinfo{person}{Eojin Lee}, \bibinfo{person}{Sangpyo Kim}, \bibinfo{person}{Keewoo Lee}, \bibinfo{person}{Namhoon Kim}, \bibinfo{person}{Chohong Min}, \bibinfo{person}{Jung~Hee Cheon}, {and} \bibinfo{person}{Jung~Ho Ahn}.} \bibinfo{year}{2020}\natexlab{}.
\newblock \bibinfo{title}{{HEAAN} Demystified: Accelerating Fully Homomorphic Encryption Through Architecture-centric Analysis and Optimization}.
\newblock \bibinfo{howpublished}{arXiv:2003.04510}.
\newblock

\bibitem[Kang et~al\mbox{.}(2013)]{kang13}
\bibfield{author}{\bibinfo{person}{Hyeong-Ju Kang}, \bibinfo{person}{Byung-Do Yang}, {and} \bibinfo{person}{Jong-Yeol Lee}.} \bibinfo{year}{2013}\natexlab{}.
\newblock \showarticletitle{Low Complexity Twiddle Factor Multiplication with {ROM} Partitioning in {FFT} Processor}.
\newblock \bibinfo{journal}{\emph{Electronics Letters}} \bibinfo{volume}{49}, \bibinfo{number}{9} (\bibinfo{year}{2013}), \bibinfo{pages}{589--591}.
\newblock
\href{https://doi.org/10.1049/el.2013.0689}{doi:\nolinkurl{10.1049/el.2013.0689}}

\bibitem[Kim et~al\mbox{.}(2024)]{cheddar24}
\bibfield{author}{\bibinfo{person}{J. Kim}, \bibinfo{person}{W. Choi}, {and} \bibinfo{person}{J. Ahn}.} \bibinfo{year}{2024}\natexlab{}.
\newblock \showarticletitle{Cheddar: A Swift Fully Homomorphic Encryption Library for CUDA GPUs}.
\newblock \bibinfo{journal}{\emph{arXiv:2407.13055}} (\bibinfo{year}{2024}).
\newblock

\bibitem[Kim et~al\mbox{.}(2023)]{sharp}
\bibfield{author}{\bibinfo{person}{Jongmin Kim}, \bibinfo{person}{Sangpyo Kim}, \bibinfo{person}{Jaewan Choi}, \bibinfo{person}{Jaiyoung Park}, \bibinfo{person}{Donghwan Kim}, {and} \bibinfo{person}{Jung~Ho Ahn}.} \bibinfo{year}{2023}\natexlab{}.
\newblock \showarticletitle{SHARP: A Short-Word Hierarchical Accelerator for Robust and Practical Fully Homomorphic Encryption}. In \bibinfo{booktitle}{\emph{50th Annual International Symposium on Computer Architecture}} \emph{(\bibinfo{series}{ISCA})}.
\newblock

\bibitem[Kim et~al\mbox{.}(2022b)]{ark}
\bibfield{author}{\bibinfo{person}{Jongmin Kim}, \bibinfo{person}{Gwangho Lee}, \bibinfo{person}{Sangpyo Kim}, \bibinfo{person}{Gina Sohn}, \bibinfo{person}{Minsoo Rhu}, \bibinfo{person}{John Kim}, {and} \bibinfo{person}{Jung~Ho Ahn}.} \bibinfo{year}{2022}\natexlab{b}.
\newblock \showarticletitle{ARK: Fully Homomorphic Encryption Accelerator with Runtime Data Generation and Inter-Operation Key Reuse}. In \bibinfo{booktitle}{\emph{IEEE/ACM International Symposium on Microarchitecture (MICRO)}}.
\newblock

\bibitem[Kim et~al\mbox{.}(2020a)]{GPU_NTT_KIM}
\bibfield{author}{\bibinfo{person}{Sangpyo Kim}, \bibinfo{person}{Wonkyung Jung}, \bibinfo{person}{Jaiyoung Park}, {and} \bibinfo{person}{Jung~Ho Ahn}.} \bibinfo{year}{2020}\natexlab{a}.
\newblock \showarticletitle{Accelerating Number Theoretic Transformations for Bootstrappable Homomorphic Encryption on GPUs}. In \bibinfo{booktitle}{\emph{2020 IEEE International Symposium on Workload Characterization (IISWC)}}. \bibinfo{pages}{264--275}.
\newblock
\href{https://doi.org/10.1109/IISWC50251.2020.00033}{doi:\nolinkurl{10.1109/IISWC50251.2020.00033}}

\bibitem[Kim et~al\mbox{.}(2022a)]{BTS}
\bibfield{author}{\bibinfo{person}{Sangpyo Kim}, \bibinfo{person}{Jongmin Kim}, \bibinfo{person}{Michael~Jaemin Kim}, \bibinfo{person}{Wonkyung Jung}, \bibinfo{person}{John Kim}, \bibinfo{person}{Minsoo Rhu}, {and} \bibinfo{person}{Jung~Ho Ahn}.} \bibinfo{year}{2022}\natexlab{a}.
\newblock \showarticletitle{BTS: An Accelerator for Bootstrappable Fully Homomorphic Encryption}. In \bibinfo{booktitle}{\emph{Proceedings of the 49th Annual International Symposium on Computer Architecture}}. \bibinfo{pages}{711–725}.
\newblock

\bibitem[Kim et~al\mbox{.}(2020b)]{kim_bug20}
\bibfield{author}{\bibinfo{person}{Sunwoong Kim}, \bibinfo{person}{Keewoo Lee}, \bibinfo{person}{Wonhee Cho}, \bibinfo{person}{Yujin Nam}, \bibinfo{person}{Jung~Hee Cheon}, {and} \bibinfo{person}{Rob~A. Rutenbar}.} \bibinfo{year}{2020}\natexlab{b}.
\newblock \showarticletitle{Hardware Architecture of a Number Theoretic Transform for a Bootstrappable {RNS}-based Homomorphic Encryption Scheme}. In \bibinfo{booktitle}{\emph{IEEE 28th Annual International Symposium on Field-Programmable Custom Computing Machines (FCCM)}}. \bibinfo{pages}{56--64}.
\newblock
\urldef\tempurl\url{https://ieeexplore.ieee.org/document/9114594}
\showURL{\tempurl}

\bibitem[Ko{\c{c}}er et~al\mbox{.}(2025)]{sevenstep}
\bibfield{author}{\bibinfo{person}{Emre Ko{\c{c}}er}, \bibinfo{person}{Selim Kirbiyik}, \bibinfo{person}{Tolun Tosun}, \bibinfo{person}{Ersin Alaybeyo{\u{g}}lu}, {and} \bibinfo{person}{Erkay Sava{\c{s}}}.} \bibinfo{year}{2025}\natexlab{}.
\newblock \showarticletitle{{IO}-Optimized Design-Time Configurable Negacyclic Seven-Step {NTT} Architecture for {FHE} Applications}. In \bibinfo{booktitle}{\emph{Proceedings of the Great Lakes Symposium on VLSI (GLSVLSI)}}.
\newblock
\href{https://doi.org/10.1145/3716368.3735514}{doi:\nolinkurl{10.1145/3716368.3735514}}

\bibitem[Ko{\c{c}}er et~al\mbox{.}(2026)]{tpntt}
\bibfield{author}{\bibinfo{person}{Emre Ko{\c{c}}er}, \bibinfo{person}{Tolun Tosun}, \bibinfo{person}{Beren Aydo{\u{g}}an}, \bibinfo{person}{Erkay Sava{\c{s}}}, \bibinfo{person}{Furkan Turan}, {and} \bibinfo{person}{Ingrid Verbauwhede}.} \bibinfo{year}{2026}\natexlab{}.
\newblock \showarticletitle{{TP-NTT}: Batch {NTT} Hardware with Application to Relinearization}.
\newblock \bibinfo{journal}{\emph{IACR Transactions on Cryptographic Hardware and Embedded Systems}} \bibinfo{volume}{2026}, \bibinfo{number}{4} (\bibinfo{year}{2026}), \bibinfo{pages}{705--730}.
\newblock
\href{https://doi.org/10.46586/tches.v2026.i4.705-730}{doi:\nolinkurl{10.46586/tches.v2026.i4.705-730}}

\bibitem[Krieger et~al\mbox{.}(2024)]{openntt}
\bibfield{author}{\bibinfo{person}{Florian Krieger}, \bibinfo{person}{Florian Hirner}, \bibinfo{person}{Ahmet~Can Mert}, {and} \bibinfo{person}{Sujoy Sinha~Roy}.} \bibinfo{year}{2024}\natexlab{}.
\newblock \showarticletitle{{OpenNTT}: An Automated Toolchain for Compiling High-Performance {NTT} Accelerators in {FHE}}. In \bibinfo{booktitle}{\emph{Proceedings of the 43rd IEEE/ACM International Conference on Computer-Aided Design (ICCAD)}}. \bibinfo{pages}{1--9}.
\newblock
\href{https://doi.org/10.1145/3676536.3697123}{doi:\nolinkurl{10.1145/3676536.3697123}}

\bibitem[Kumarathunga et~al\mbox{.}(2025)]{autontt}
\bibfield{author}{\bibinfo{person}{Dilshan Kumarathunga}, \bibinfo{person}{Qilin Hu}, {and} \bibinfo{person}{Zhenman Fang}.} \bibinfo{year}{2025}\natexlab{}.
\newblock \showarticletitle{AutoNTT: Automatic Architecture Design and Exploration for Number Theoretic Transform Acceleration on FPGAs}. In \bibinfo{booktitle}{\emph{2025 IEEE 33rd Annual International Symposium on Field-Programmable Custom Computing Machines (FCCM)}}. \bibinfo{pages}{1--9}.
\newblock
\href{https://doi.org/10.1109/FCCM62733.2025.00024}{doi:\nolinkurl{10.1109/FCCM62733.2025.00024}}

\bibitem[Kundi et~al\mbox{.}(2024)]{kundi_ashes24}
\bibfield{author}{\bibinfo{person}{Dur E~Shahwar Kundi}, \bibinfo{person}{Jose~M. Bermudo~Mera}, \bibinfo{person}{Pierre-Yves Strub}, {and} \bibinfo{person}{Michael Hutter}.} \bibinfo{year}{2024}\natexlab{}.
\newblock \showarticletitle{High-Performance {NTT} Hardware Accelerator to Support {ML-KEM} and {ML-DSA}}. In \bibinfo{booktitle}{\emph{Proceedings of the 2024 Workshop on Attacks and Solutions in Hardware Security (ASHES)}}. \bibinfo{pages}{100--105}.
\newblock
\href{https://doi.org/10.1145/3689939.3695785}{doi:\nolinkurl{10.1145/3689939.3695785}}

\bibitem[Kurniawan et~al\mbox{.}(2023)]{config-ntt}
\bibfield{author}{\bibinfo{person}{Stefanus Kurniawan}, \bibinfo{person}{Phap Duong-Ngoc}, {and} \bibinfo{person}{Hanho Lee}.} \bibinfo{year}{2023}\natexlab{}.
\newblock \showarticletitle{Configurable Memory-Based NTT Architecture for Homomorphic Encryption}.
\newblock \bibinfo{journal}{\emph{IEEE Transactions on Circuits and Systems II: Express Briefs}} \bibinfo{volume}{70}, \bibinfo{number}{10} (\bibinfo{year}{2023}), \bibinfo{pages}{3942--3946}.
\newblock
\href{https://doi.org/10.1109/TCSII.2023.3289489}{doi:\nolinkurl{10.1109/TCSII.2023.3289489}}

\bibitem[Lee et~al\mbox{.}(2024)]{config-ntt-arithmatic-core}
\bibfield{author}{\bibinfo{person}{Chulwoo Lee}, \bibinfo{person}{Hanyoung Lee}, \bibinfo{person}{Ardianto Satriawan}, {and} \bibinfo{person}{Hanho Lee}.} \bibinfo{year}{2024}\natexlab{}.
\newblock \showarticletitle{Configurable Arithmetic Core Architecture for RNS-CKKS Homomorphic Encryption}.
\newblock \bibinfo{journal}{\emph{IEEE Access}}  \bibinfo{volume}{PP} (\bibinfo{date}{01} \bibinfo{year}{2024}), \bibinfo{pages}{1--1}.
\newblock
\href{https://doi.org/10.1109/ACCESS.2024.3473903}{doi:\nolinkurl{10.1109/ACCESS.2024.3473903}}

\bibitem[Lee et~al\mbox{.}(2026)]{lee_access26}
\bibfield{author}{\bibinfo{person}{Ju-Hyung Lee}, \bibinfo{person}{Young-Chan Jang}, {and} \bibinfo{person}{Kyung-Wook Shin}.} \bibinfo{year}{2026}\natexlab{}.
\newblock \showarticletitle{A Radix-2/4/8-Based Pipelined {NTT/INTT} Architecture for Homomorphic Encryption}.
\newblock \bibinfo{journal}{\emph{IEEE Access}}  \bibinfo{volume}{14} (\bibinfo{year}{2026}), \bibinfo{pages}{87042--87058}.
\newblock
\href{https://doi.org/10.1109/ACCESS.2026.3699867}{doi:\nolinkurl{10.1109/ACCESS.2026.3699867}}

\bibitem[Liu et~al\mbox{.}(2024)]{hmntt}
\bibfield{author}{\bibinfo{person}{Chen Liu} {et~al\mbox{.}}} \bibinfo{year}{2024}\natexlab{}.
\newblock \showarticletitle{{HMNTT}: A Highly Efficient {MDC-NTT} Architecture for Privacy-preserving Applications}. In \bibinfo{booktitle}{\emph{Proceedings of the Great Lakes Symposium on VLSI (GLSVLSI)}}.
\newblock
\href{https://doi.org/10.1145/3649476.3658734}{doi:\nolinkurl{10.1145/3649476.3658734}}

\bibitem[Lyubashevsky et~al\mbox{.}(2010)]{EC:LyuPeiReg10}
\bibfield{author}{\bibinfo{person}{Vadim Lyubashevsky}, \bibinfo{person}{Chris Peikert}, {and} \bibinfo{person}{Oded Regev}.} \bibinfo{year}{2010}\natexlab{}.
\newblock \showarticletitle{On Ideal Lattices and Learning with Errors over Rings}. In \bibinfo{booktitle}{\emph{Advances in Cryptology – EUROCRYPT 2010}} \emph{(\bibinfo{series}{Lecture Notes in Computer Science}, Vol.~\bibinfo{volume}{6110})}. \bibinfo{publisher}{Springer}, \bibinfo{pages}{1--23}.
\newblock
\href{https://doi.org/10.1007/978-3-642-13190-5_1}{doi:\nolinkurl{10.1007/978-3-642-13190-5_1}}

\bibitem[Mareta et~al\mbox{.}(2024)]{bootstrapping-capable-ntt}
\bibfield{author}{\bibinfo{person}{Rella Mareta}, \bibinfo{person}{Ardianto Satriawan}, \bibinfo{person}{Phap~Ngoc Duong}, {and} \bibinfo{person}{Hanho Lee}.} \bibinfo{year}{2024}\natexlab{}.
\newblock \showarticletitle{A Bootstrapping-Capable Configurable NTT Architecture for Fully Homomorphic Encryption}.
\newblock \bibinfo{journal}{\emph{IEEE Access}}  \bibinfo{volume}{12} (\bibinfo{year}{2024}), \bibinfo{pages}{52911--52921}.
\newblock
\href{https://doi.org/10.1109/ACCESS.2024.3386977}{doi:\nolinkurl{10.1109/ACCESS.2024.3386977}}

\bibitem[Mert et~al\mbox{.}(2023)]{medha}
\bibfield{author}{\bibinfo{person}{Ahmet~Can Mert}, \bibinfo{person}{{Aikata}}, \bibinfo{person}{Sunmin Kwon}, \bibinfo{person}{Youngsam Shin}, \bibinfo{person}{Donghoon Yoo}, \bibinfo{person}{Yongwoo Lee}, {and} \bibinfo{person}{Sujoy Sinha~Roy}.} \bibinfo{year}{2023}\natexlab{}.
\newblock \showarticletitle{{Medha}: Microcoded Hardware Accelerator for computing on Encrypted Data}.
\newblock \bibinfo{journal}{\emph{IACR Transactions on Cryptographic Hardware and Embedded Systems}} \bibinfo{volume}{2023}, \bibinfo{number}{1} (\bibinfo{year}{2023}), \bibinfo{pages}{463--500}.
\newblock
\href{https://doi.org/10.46586/tches.v2023.i1.463-500}{doi:\nolinkurl{10.46586/tches.v2023.i1.463-500}}

\bibitem[Ozerk et~al\mbox{.}(2021)]{GPU_NTT_Ozerk}
\bibfield{author}{\bibinfo{person}{Ozgun Ozerk}, \bibinfo{person}{Can Elgezen}, \bibinfo{person}{Ahmet~Can Mert}, \bibinfo{person}{Erdinc Ozturk}, {and} \bibinfo{person}{Erkay Savas}.} \bibinfo{year}{2021}\natexlab{}.
\newblock \bibinfo{title}{Efficient Number Theoretic Transform Implementation on {GPU} for Homomorphic Encryption}.
\newblock \bibinfo{howpublished}{Cryptology {ePrint} Archive, Paper 2021/124}.
\newblock
\urldef\tempurl\url{https://eprint.iacr.org/2021/124}
\showURL{\tempurl}

\bibitem[Pindado et~al\mbox{.}(2026)]{survey_home}
\bibfield{author}{\bibinfo{person}{Zaira Pindado}, \bibinfo{person}{Thomas Spendlhofer}, \bibinfo{person}{Mohamed Allam}, \bibinfo{person}{Priyam Mehta}, \bibinfo{person}{Lena Martens}, {and} \bibinfo{person}{Antonio~J. Peña}.} \bibinfo{year}{2026}\natexlab{}.
\newblock \bibinfo{title}{{SoK} of Private Deep Neural Network Inference with Approximate Fully Homomorphic Encryption}.
\newblock \bibinfo{howpublished}{Cryptology {ePrint} Archive, Paper 2026/047}.
\newblock
\urldef\tempurl\url{https://eprint.iacr.org/2026/047}
\showURL{\tempurl}

\bibitem[Regev(2009)]{regev2009lattices}
\bibfield{author}{\bibinfo{person}{Oded Regev}.} \bibinfo{year}{2009}\natexlab{}.
\newblock \showarticletitle{On Lattices, Learning with Errors, Random Linear Codes, and Cryptography}.
\newblock \bibinfo{journal}{\emph{Journal of the ACM (JACM)}} \bibinfo{volume}{56}, \bibinfo{number}{6} (\bibinfo{year}{2009}), \bibinfo{pages}{1--40}.
\newblock

\bibitem[Riazi et~al\mbox{.}(2020)]{heax}
\bibfield{author}{\bibinfo{person}{M.~Sadegh Riazi}, \bibinfo{person}{Kim Laine}, \bibinfo{person}{Blake Pelton}, {and} \bibinfo{person}{Wei Dai}.} \bibinfo{year}{2020}\natexlab{}.
\newblock \showarticletitle{HEAX: An Architecture for Computing on Encrypted Data}. In \bibinfo{booktitle}{\emph{Proceedings of the 25th International Conference on Architectural Support for Programming Languages and Operating Systems}} \emph{(\bibinfo{series}{ASPLOS '20})}. \bibinfo{publisher}{Association for Computing Machinery}, \bibinfo{pages}{1295--1309}.
\newblock
\href{https://doi.org/10.1145/3373376.3378523}{doi:\nolinkurl{10.1145/3373376.3378523}}

\bibitem[Samardzic et~al\mbox{.}(2021)]{F12021}
\bibfield{author}{\bibinfo{person}{Nikola Samardzic}, \bibinfo{person}{Axel Feldmann}, \bibinfo{person}{Aleksandar Krastev}, \bibinfo{person}{Srinivas Devadas}, \bibinfo{person}{Ronald Dreslinski}, \bibinfo{person}{Christopher Peikert}, {and} \bibinfo{person}{Daniel Sanchez}.} \bibinfo{year}{2021}\natexlab{}.
\newblock \showarticletitle{F1: A Fast and Programmable Accelerator for Fully Homomorphic Encryption}. In \bibinfo{booktitle}{\emph{MICRO-54: 54th Annual IEEE/ACM International Symposium on Microarchitecture}} (Virtual Event, Greece) \emph{(\bibinfo{series}{MICRO '21})}. \bibinfo{publisher}{Association for Computing Machinery}, \bibinfo{address}{New York, NY, USA}, \bibinfo{pages}{238–252}.
\newblock
\showISBNx{9781450385572}
\href{https://doi.org/10.1145/3466752.3480070}{doi:\nolinkurl{10.1145/3466752.3480070}}

\bibitem[Samardzic et~al\mbox{.}(2022)]{craterlake}
\bibfield{author}{\bibinfo{person}{Nikola Samardzic}, \bibinfo{person}{Axel Feldmann}, \bibinfo{person}{Aleksandar Krastev}, \bibinfo{person}{Nathan Manohar}, \bibinfo{person}{Nicholas Genise}, \bibinfo{person}{Srinivas Devadas}, \bibinfo{person}{Karim Eldefrawy}, \bibinfo{person}{Chris Peikert}, {and} \bibinfo{person}{Daniel Sanchez}.} \bibinfo{year}{2022}\natexlab{}.
\newblock \showarticletitle{CraterLake: A Hardware Accelerator for Efficient Unbounded Computation on Encrypted Data}. In \bibinfo{booktitle}{\emph{Int. Symp. Comp. Arch.}}
\newblock

\bibitem[Shivdikar et~al\mbox{.}(2022)]{Shivdikar2022}
\bibfield{author}{\bibinfo{person}{Kaustubh Shivdikar}, \bibinfo{person}{Gilbert Jonatan}, \bibinfo{person}{Evelio Mora}, \bibinfo{person}{Neal Livesay}, \bibinfo{person}{Rashmi Agrawal}, \bibinfo{person}{Ajay Joshi}, \bibinfo{person}{Jose~L. Abellan}, \bibinfo{person}{John Kim}, {and} \bibinfo{person}{David Kaeli}.} \bibinfo{year}{2022}\natexlab{}.
\newblock \showarticletitle{{ Accelerating Polynomial Multiplication for Homomorphic Encryption on GPUs }}. In \bibinfo{booktitle}{\emph{2022 IEEE International Symposium on Secure and Private Execution Environment Design (SEED)}}. \bibinfo{publisher}{IEEE Computer Society}, \bibinfo{address}{Los Alamitos, CA, USA}, \bibinfo{pages}{61--72}.
\newblock
\href{https://doi.org/10.1109/SEED55351.2022.00013}{doi:\nolinkurl{10.1109/SEED55351.2022.00013}}

\bibitem[Stelzer et~al\mbox{.}(2026)]{stelzer_tcas26}
\bibfield{author}{\bibinfo{person}{Tobias Stelzer}, \bibinfo{person}{Patrick Karl}, {and} \bibinfo{person}{Andreas Zankl}.} \bibinfo{year}{2026}\natexlab{}.
\newblock \showarticletitle{Flexible Parallel Radix-4 {MDC} {NTT} for {FHE}}.
\newblock \bibinfo{journal}{\emph{IEEE Transactions on Circuits and Systems II: Express Briefs}} \bibinfo{volume}{73}, \bibinfo{number}{8} (\bibinfo{year}{2026}), \bibinfo{pages}{863--867}.
\newblock
\href{https://doi.org/10.1109/TCSII.2026.3700928}{doi:\nolinkurl{10.1109/TCSII.2026.3700928}}

\bibitem[Takala et~al\mbox{.}(2000)]{CG}
\bibfield{author}{\bibinfo{person}{J. Takala}, \bibinfo{person}{D. Akopian}, \bibinfo{person}{J. Astola}, {and} \bibinfo{person}{J. Saarinen}.} \bibinfo{year}{2000}\natexlab{}.
\newblock \showarticletitle{Constant Geometry Algorithm for Discrete Cosine Transform}.
\newblock \bibinfo{journal}{\emph{IEEE Transactions on Signal Processing}} \bibinfo{volume}{48}, \bibinfo{number}{6} (\bibinfo{year}{2000}), \bibinfo{pages}{1840--1843}.
\newblock
\href{https://doi.org/10.1109/78.845951}{doi:\nolinkurl{10.1109/78.845951}}

\bibitem[Tan et~al\mbox{.}(2024)]{parentt}
\bibfield{author}{\bibinfo{person}{Weihang Tan}, \bibinfo{person}{Sin-Wei Chiu}, \bibinfo{person}{Antian Wang}, \bibinfo{person}{Yingjie Lao}, {and} \bibinfo{person}{Keshab~K. Parhi}.} \bibinfo{year}{2024}\natexlab{}.
\newblock \showarticletitle{{PaReNTT}: Low-Latency Parallel Residue Number System and {NTT}-Based Long Polynomial Modular Multiplication for Homomorphic Encryption}.
\newblock \bibinfo{journal}{\emph{IEEE Transactions on Information Forensics and Security}}  \bibinfo{volume}{19} (\bibinfo{year}{2024}), \bibinfo{pages}{1646--1659}.
\newblock
\href{https://doi.org/10.1109/TIFS.2023.3338553}{doi:\nolinkurl{10.1109/TIFS.2023.3338553}}

\bibitem[Wang and Gao(2023)]{sam}
\bibfield{author}{\bibinfo{person}{Cheng Wang} {and} \bibinfo{person}{Mingyu Gao}.} \bibinfo{year}{2023}\natexlab{}.
\newblock \showarticletitle{SAM: A Scalable Accelerator for Number Theoretic Transform Using Multi-Dimensional Decomposition}. In \bibinfo{booktitle}{\emph{2023 IEEE/ACM International Conference on Computer Aided Design (ICCAD)}}. \bibinfo{pages}{1--9}.
\newblock
\href{https://doi.org/10.1109/ICCAD57390.2023.10323744}{doi:\nolinkurl{10.1109/ICCAD57390.2023.10323744}}

\bibitem[Wang and Ikeda(2024)]{wang_ikeda_elex24}
\bibfield{author}{\bibinfo{person}{Zeyu Wang} {and} \bibinfo{person}{Makoto Ikeda}.} \bibinfo{year}{2024}\natexlab{}.
\newblock \showarticletitle{High-Throughput and Fully-Pipelined Ciphertext Multiplier for Homomorphic Encryption}.
\newblock \bibinfo{journal}{\emph{IEICE Electronics Express}} \bibinfo{volume}{21}, \bibinfo{number}{6} (\bibinfo{year}{2024}), \bibinfo{pages}{20230628}.
\newblock
\href{https://doi.org/10.1587/elex.21.20230628}{doi:\nolinkurl{10.1587/elex.21.20230628}}

\bibitem[Wold and Despain(1984)]{wold_despain84}
\bibfield{author}{\bibinfo{person}{Erling~H. Wold} {and} \bibinfo{person}{Alvin~M. Despain}.} \bibinfo{year}{1984}\natexlab{}.
\newblock \showarticletitle{Pipeline and Parallel-Pipeline {FFT} Processors for {VLSI} Implementations}.
\newblock \bibinfo{journal}{\emph{IEEE Trans. Comput.}} \bibinfo{volume}{C-33}, \bibinfo{number}{5} (\bibinfo{year}{1984}), \bibinfo{pages}{414--426}.
\newblock

\bibitem[Xu et~al\mbox{.}(2026)]{hera}
\bibfield{author}{\bibinfo{person}{Zhihan Xu}, \bibinfo{person}{Rajgopal Kannan}, {and} \bibinfo{person}{Viktor~K. Prasanna}.} \bibinfo{year}{2026}\natexlab{}.
\newblock \showarticletitle{{HERA}: A Bandwidth-efficient Accelerator for Fully Homomorphic Encryption on {HBM}-enabled {FPGA}}. In \bibinfo{booktitle}{\emph{Proceedings of the ACM/SIGDA International Symposium on Field Programmable Gate Arrays (FPGA)}}. \bibinfo{pages}{265--276}.
\newblock
\href{https://doi.org/10.1145/3748173.3779201}{doi:\nolinkurl{10.1145/3748173.3779201}}

\bibitem[Xu et~al\mbox{.}(2025)]{fast}
\bibfield{author}{\bibinfo{person}{Zhihan Xu}, \bibinfo{person}{Tian Ye}, \bibinfo{person}{Rajgopal Kannan}, {and} \bibinfo{person}{Viktor~K. Prasanna}.} \bibinfo{year}{2025}\natexlab{}.
\newblock \showarticletitle{FAST: FPGA Acceleration of Fully Homomorphic Encryption with Efficient Bootstrapping}. In \bibinfo{booktitle}{\emph{Proceedings of the 2025 ACM/SIGDA International Symposium on Field Programmable Gate Arrays}} \emph{(\bibinfo{series}{FPGA '25})}. \bibinfo{publisher}{Association for Computing Machinery}, \bibinfo{address}{New York, NY, USA}, \bibinfo{pages}{115–126}.
\newblock
\href{https://doi.org/10.1145/3706628.3708879}{doi:\nolinkurl{10.1145/3706628.3708879}}

\bibitem[Yang et~al\mbox{.}(2022)]{nttgen}
\bibfield{author}{\bibinfo{person}{Yang Yang}, \bibinfo{person}{Sanmukh~R. Kuppannagari}, \bibinfo{person}{Rajgopal Kannan}, {and} \bibinfo{person}{Viktor~K. Prasanna}.} \bibinfo{year}{2022}\natexlab{}.
\newblock \showarticletitle{NTTGen: A Framework for Generating Low Latency NTT Implementations on FPGA}. In \bibinfo{booktitle}{\emph{Proceedings of the 19th ACM International Conference on Computing Frontiers}} (Turin, Italy) \emph{(\bibinfo{series}{CF '22})}. \bibinfo{publisher}{Association for Computing Machinery}, \bibinfo{address}{New York, NY, USA}, \bibinfo{pages}{30–39}.
\newblock
\showISBNx{9781450393386}
\href{https://doi.org/10.1145/3528416.3530225}{doi:\nolinkurl{10.1145/3528416.3530225}}

\bibitem[Yang et~al\mbox{.}(2023)]{poseidon}
\bibfield{author}{\bibinfo{person}{Yinghao Yang}, \bibinfo{person}{Huaizhi Zhang}, \bibinfo{person}{Shengyu Fan}, \bibinfo{person}{Hang Lu}, \bibinfo{person}{Mingzhe Zhang}, {and} \bibinfo{person}{Xiaowei Li}.} \bibinfo{year}{2023}\natexlab{}.
\newblock \showarticletitle{Poseidon: Practical Homomorphic Encryption Accelerator}. In \bibinfo{booktitle}{\emph{2023 IEEE International Symposium on High-Performance Computer Architecture (HPCA)}}. \bibinfo{pages}{870--881}.
\newblock
\href{https://doi.org/10.1109/HPCA56546.2023.10070984}{doi:\nolinkurl{10.1109/HPCA56546.2023.10070984}}

\bibitem[Ye et~al\mbox{.}(2021)]{ye_isc21}
\bibfield{author}{\bibinfo{person}{Tian Ye}, \bibinfo{person}{Yang Yang}, \bibinfo{person}{Sanmukh~R. Kuppannagari}, \bibinfo{person}{Rajgopal Kannan}, {and} \bibinfo{person}{Viktor~K. Prasanna}.} \bibinfo{year}{2021}\natexlab{}.
\newblock \showarticletitle{{FPGA} Acceleration of Number Theoretic Transform}. In \bibinfo{booktitle}{\emph{High Performance Computing (ISC High Performance)}} \emph{(\bibinfo{series}{Lecture Notes in Computer Science})}. \bibinfo{publisher}{Springer}.
\newblock
\href{https://doi.org/10.1007/978-3-030-78713-4_6}{doi:\nolinkurl{10.1007/978-3-030-78713-4_6}}

\bibitem[Ye et~al\mbox{.}(2022)]{pipentt}
\bibfield{author}{\bibinfo{person}{Zhaoxiang Ye}, \bibinfo{person}{Ray C.~C. Cheung}, {and} \bibinfo{person}{Kai Huang}.} \bibinfo{year}{2022}\natexlab{}.
\newblock \showarticletitle{{PipeNTT}: A Pipelined Number Theoretic Transform Architecture}.
\newblock \bibinfo{journal}{\emph{IEEE Transactions on Circuits and Systems II: Express Briefs}} \bibinfo{volume}{69}, \bibinfo{number}{10} (\bibinfo{year}{2022}), \bibinfo{pages}{4068--4072}.
\newblock

\bibitem[Yune et~al\mbox{.}(2025)]{abcfhe}
\bibfield{author}{\bibinfo{person}{Sungwoong Yune}, \bibinfo{person}{Hyojeong Lee}, \bibinfo{person}{Adiwena Putra}, \bibinfo{person}{Hyunjun Cho}, \bibinfo{person}{Cuong Duong~Manh}, \bibinfo{person}{Jaeho Jeon}, {and} \bibinfo{person}{Joo-Young Kim}.} \bibinfo{year}{2025}\natexlab{}.
\newblock \showarticletitle{{ABC-FHE}: A Resource-Efficient Accelerator Enabling Bootstrappable Parameters for Client-Side Fully Homomorphic Encryption}. In \bibinfo{booktitle}{\emph{62nd ACM/IEEE Design Automation Conference (DAC)}}.
\newblock

\bibitem[Zaman et~al\mbox{.}(2026)]{zaman_tcas26}
\bibfield{author}{\bibinfo{person}{Moonis~Amir Zaman}, \bibinfo{person}{Muhammad~Bilal Khan}, \bibinfo{person}{Waqar Ahmad}, \bibinfo{person}{Ayesha Khalid}, {and} \bibinfo{person}{M{\'a}ire O'Neill}.} \bibinfo{year}{2026}\natexlab{}.
\newblock \showarticletitle{A Lightweight and Hardware-Efficient {NTT} {FPGA} Accelerator for {FHE} Applications}.
\newblock \bibinfo{journal}{\emph{IEEE Transactions on Circuits and Systems II: Express Briefs}} \bibinfo{volume}{73}, \bibinfo{number}{1} (\bibinfo{year}{2026}), \bibinfo{pages}{78--82}.
\newblock
\href{https://doi.org/10.1109/TCSII.2025.3633181}{doi:\nolinkurl{10.1109/TCSII.2025.3633181}}

\bibitem[Zhai et~al\mbox{.}(2022)]{GPU_NTT_ZHAI}
\bibfield{author}{\bibinfo{person}{Yujia Zhai}, \bibinfo{person}{Mohannad Ibrahim}, \bibinfo{person}{Yiqin Qiu}, \bibinfo{person}{Fabian Boemer}, \bibinfo{person}{Zizhong Chen}, \bibinfo{person}{Alexey Titov}, {and} \bibinfo{person}{Alexander Lyashevsky}.} \bibinfo{year}{2022}\natexlab{}.
\newblock \showarticletitle{Accelerating Encrypted Computing on Intel GPUs}. In \bibinfo{booktitle}{\emph{2022 IEEE International Parallel and Distributed Processing Symposium (IPDPS)}}. \bibinfo{pages}{705--716}.
\newblock
\href{https://doi.org/10.1109/IPDPS53621.2022.00074}{doi:\nolinkurl{10.1109/IPDPS53621.2022.00074}}

\bibitem[Zhang et~al\mbox{.}(2026)]{bsr_tcad26}
\bibfield{author}{\bibinfo{person}{Bo Zhang}, \bibinfo{person}{Mingzhe Zhang}, {and} \bibinfo{person}{Shoumeng Yan}.} \bibinfo{year}{2026}\natexlab{}.
\newblock \showarticletitle{Binary Subscript Representation and Efficient Pipelined Number Theoretic Transform}.
\newblock \bibinfo{journal}{\emph{IEEE Transactions on Computer-Aided Design of Integrated Circuits and Systems}} \bibinfo{volume}{45}, \bibinfo{number}{9} (\bibinfo{year}{2026}), \bibinfo{pages}{4316--4329}.
\newblock
\href{https://doi.org/10.1109/TCAD.2025.3648669}{doi:\nolinkurl{10.1109/TCAD.2025.3648669}}

\bibitem[Zhang et~al\mbox{.}(2021)]{pipezk}
\bibfield{author}{\bibinfo{person}{Ye Zhang}, \bibinfo{person}{Shuo Wang}, \bibinfo{person}{Xian Zhang}, \bibinfo{person}{Jiangbin Dong}, \bibinfo{person}{Xingzhong Mao}, \bibinfo{person}{Fan Long}, \bibinfo{person}{Cong Wang}, \bibinfo{person}{Dong Zhou}, \bibinfo{person}{Mingyu Gao}, {and} \bibinfo{person}{Guangyu Sun}.} \bibinfo{year}{2021}\natexlab{}.
\newblock \showarticletitle{{PipeZK}: Accelerating Zero-Knowledge Proof with a Pipelined Architecture}. In \bibinfo{booktitle}{\emph{ACM/IEEE 48th Annual International Symposium on Computer Architecture (ISCA)}}.
\newblock

\bibitem[Zhu et~al\mbox{.}(2026)]{autonest}
\bibfield{author}{\bibinfo{person}{Yilan Zhu}, \bibinfo{person}{Geng Yang}, \bibinfo{person}{Xingyu Tian}, \bibinfo{person}{Dilshan Kumarathunga}, \bibinfo{person}{Liang Kong}, \bibinfo{person}{Xianglong Deng}, \bibinfo{person}{Shengyu Fan}, \bibinfo{person}{Guang Fan}, \bibinfo{person}{Guiming Shi}, \bibinfo{person}{Lei Chen}, \bibinfo{person}{Bo Zhang}, \bibinfo{person}{Yisong Chang}, \bibinfo{person}{Shoumeng Yan}, \bibinfo{person}{Zhenman Fang}, {and} \bibinfo{person}{Mingzhe Zhang}.} \bibinfo{year}{2026}\natexlab{}.
\newblock \showarticletitle{An Efficient and Scalable Hardware Architecture for Number Theoretic Transform on FPGA with Design Automation}. In \bibinfo{booktitle}{\emph{2026 IEEE International Symposium on High Performance Computer Architecture (HPCA)}}.
\newblock
\href{https://doi.org/10.1109/HPCA68181.2026.11408615}{doi:\nolinkurl{10.1109/HPCA68181.2026.11408615}}

\bibitem[Özcan and Savaş(2023)]{Ozcan2023}
\bibfield{author}{\bibinfo{person}{Ege Özcan} {and} \bibinfo{person}{Erkay Savaş}.} \bibinfo{year}{2023}\natexlab{}.
\newblock \bibinfo{title}{Two Algorithms for Fast GPU Implementation of the Number Theoretic Transform}.
\newblock \bibinfo{howpublished}{Cryptology ePrint Archive, Report 2023/1410}.
\newblock
\urldef\tempurl\url{https://eprint.iacr.org/2023/1410}
\showURL{\tempurl}

\bibitem[Özcan et~al\mbox{.}(2022)]{Ozcan2022}
\bibfield{author}{\bibinfo{person}{Ege Özcan}, \bibinfo{person}{Erkay Savaş}, {et~al\mbox{.}}} \bibinfo{year}{2022}\natexlab{}.
\newblock \bibinfo{title}{Homomorphic Encryption on GPU: Optimized NTT Kernels for BFV}.
\newblock \bibinfo{howpublished}{Cryptology ePrint Archive, Report 2022/1222}.
\newblock
\urldef\tempurl\url{https://eprint.iacr.org/2022/1222}
\showURL{\tempurl}

\end{thebibliography}

\end{document}